\documentclass[10pt]{iopart}
\usepackage[dvips]{color}
\usepackage{Graphen}
\usepackage{graphicx,pstricks,pst-plot}
\newcmykcolor{fgColor}{0.00000 0.00000 0.00000 1.00000}
\newcmykcolor{bgColor}{0.00000 0.00000 0.00000 0.00000}

\newcommand{\mathi}{{\rm i}}
\newcommand{\sgn}{{\rm sgn}}

\begin{document}
\title[Optical excitations in the one-dimensional 
extended Peierls--Hubbard model]{Perturbation theory for optical excitations
in the one-dimensional extended Peierls--Hubbard model}

\author{Anja Grage, Florian 
Gebhard 
and J\"org Rissler}

\address{Fachbereich Physik, Philipps-Universit\"at Marburg,
D-35032 Marburg, Germany}

\begin{abstract}
For the one-dimensional, extended Peierls--Hubbard model we
cal\-culate analytically the ground-state energy and the single-particle gap
to second order in the Coulomb interaction for a given lattice
dimerization. The comparison with numerically exact data from the
Density-Matrix Renormalization Group shows that
the ground-state energy is quantitatively reliable 
for Coulomb parameters as large as the band width.
The single-particle gap can almost triple from its bare Peierls
value before substantial deviations appear.
For the calculation of the dominant optical excitations, we follow
two approaches. In Wannier theory, we perturb the Wannier exciton
states to second order. In two-step perturbation theory,
similar in spirit to the GW-BSE approach,
we form excitons from dressed electron-hole excitations.
We find the Wannier approach to be superior to the two-step
perturbation theory. For singlet excitons, Wannier theory is applicable
up to Coulomb parameters as large as half band width. For triplet excitons,
second-order perturbation theory quickly fails completely.
\end{abstract}




\section{Introduction}
\label{intro}

Electrons in low-dimensional systems 
behave very differently from their three-dimensional counterparts.
On the one hand, at low temperatures they tend to distort the lattice  
(Peierls effect~\cite{Peierls}). For example, poly-acetylene as
the simplest $\pi$-conjugated polymer shows an alternation
between short and long bonds~\cite{Farges}. At the Peierls tran\-sition
a gap opens in the single-particle spectrum and the metal turns 
into a Peierls insulator.
On the other hand, the electron-electron interaction provides a relevant
perturbation in one dimension. For example, metals in
one dimension are Luttinger liquids in which
the low-energy charge and spin excitations propagate with different velocities
so that quasi-particles as in Landau Fermi-Liquid Theory are absent
in the single-particle spectral function~\cite{Voit,Giamarchi}. 
At commensurate band fillings, the electron-electron interaction induces
Umklapp scatterings which turn the metal state into a Mott insulator.

This immediately raises the question: which of the two
perturbations, Peierls distortion or Mott transition, dominates
the other one? This question can be answered within the framework of
field theory because both the electron-lattice and the
electron-electron interactions destroy the metallic phase even in the
limit of weak coupling. As has been shown in~\cite{AnjaPhD,Eckern}, the
Peierls distortion is always a relevant perturbation in the sense
of a renormalization-group analysis whereas the Umklapp scatterings
which cause the Mott transition are only marginally relevant.
At large Coulomb interactions when the system is a Mott insulator,
the Peierls distortion is also present at low temperatures
because the effective spin system undergoes a spin-Peierls transition.
Therefore, one-dimensional lattices are expected to be
Peierls distorted at zero temperature,
irrespective of the presence of the electron-electron interaction. 

In presence of a Peierls distortion and one electron per lattice site
(half band-filling),
the Peierls insulator can be described in terms of a filled
lower Peierls band and an empty upper Peierls band which are 
separated by a finite Peierls gap whose size is proportional to 
the dimerization strength. In principle, the finite Peierls gap 
permits a perturbative treatment of the electron-electron interaction.
It should be clear, however, that such a perturbative treatment
is limited to `weak coupling', and the range of the validity
of perturbation theory remains to be determined.
In particular, even the screened electron-electron interaction
in polymers is quite sizable~\cite{Kiess}, and it remains to be
studied in how far
Wannier theory and similar perturbative approaches~\cite{Abe,GWBSE} lead to
a reliable description of optical excitations in these materials.

In this work, we investigate the extended Peierls--Hubbard model
at half band-filling which models the Peierls distortion and the 
Coulomb interaction in ideal poly-acetylene. The model
is introduced in section~\ref{sec:Hamilt}. 
In section~\ref{sec:gsener} we use standard Ray\-leigh\--Schr\"odinger
perturbation theory to calculate the ground-state energy to
second order in the Coulomb interaction and compare our results
to numerically exact data from the Density-Matrix Renormalization Group
(DMRG). We show that second-order perturbation theory is
applicable up to interaction strengths as large as
the band width. In section~\ref{sec:singleP} we observe the same
behavior for the gap for single-particle excitations which increases
quickly as a function of the interaction strength.

In order to study optical excitations, we follow two routes because
Rayleigh--Schr\"odinger perturbation theory is inapplicable for a
description of bound electron-hole pairs. 
In section~\ref{sec:ptapproachestoexc} we develop Wannier perturbation 
theory to second order where an exciton is formed already in first order.
Alternatively, we introduce the `two-step
perturbation theory' in which the exciton is formed in a second step
after the electrons and holes have been dressed first.
In section~\ref{sec:excitons2nd} the comparison of our analytical
results with those from the DMRG show that second-order Wannier theory
is superior to the two-step perturbation theory. The singlet exciton
is reasonably well described by Wannier theory for moderate
interactions but perturbation theory quickly fails for the triplet
exciton. We close our presentation with
concluding remarks in section~\ref{sec:discussion}.
Technical details of the calculations are deferred to the appendix.

\section{Hamilton Operator}
\label{sec:Hamilt}

\subsection{Peierls Model}
\label{subsec:PM}

We investigate spin-1/2 electrons on a 
periodically distorted chain of $L$~sites whose motion is described by
the Peierls Hamiltonian,
\begin{equation}
\hat{T} = (-t)
\sum_{l;\sigma} \left(1+(-1)^l\delta \right)
\left(\hat{c}_{l,\sigma}^+\hat{c}_{l+1,\sigma}^{\phantom{+}}+
\hat{c}_{l+1,\sigma}^+\hat{c}_{l,\sigma}^{\phantom{+}}\right) \; ,
\end{equation}
where $\hat{c}^+_{l,\sigma}$,
$\hat{c}_{l,\sigma}$ are creation and annihilation operators for
electrons with spin~$\sigma=\uparrow,\downarrow$ on site~$l$.
The lattice spacing is set to unity, $a_0\equiv 1$.
Since we are interested
in the insulating phase, we consider exclusively a half-filled band
where the number of electrons~$N=2 N_{\sigma}$ equals the
number of lattice sites~$L$. 
For our analytical calculations, we choose periodic boundary
conditions and $L/2$ even. For the DMRG investigations,
open boundary conditions are employed.

The Peierls operator is diagonal in momentum space~\cite{Bott1},
\begin{equation}
\hat{T} = \sum_{k;\sigma} E(k) \left(
\hat{b}^+_{k,\sigma}\hat{b}_{k,\sigma}^{\phantom{+}}
-
\hat{a}^+_{k,\sigma}\hat{a}_{k,\sigma}^{\phantom{+}}\right) \; , 
\end{equation}
where $k=2\pi m/L$, $m=-L/2, \ldots, L/2-1$, are the momenta from
the reduced Brillouin zone.
The dispersion relation for the upper and lower Peierls bands is 
\begin{equation}
E(k) = \sqrt{\epsilon(k)^2+\Delta(k)^2}
\end{equation}
with
\begin{eqnarray}
\epsilon(k) &=& -2t\cos(k) \; ,\\
\Delta(k) &=& 2t\delta \sin(k) \nonumber \; .
\end{eqnarray}
The Fermi operators for the electrons in the Peierls bands obey
\begin{eqnarray}
\hat{a}_{k, \sigma} & \equiv  
\alpha_k\hat{c}_{k,\sigma}+\mathi \beta_k\hat{c}_{k+\pi,\sigma}\;,
\nonumber\\
\hat{b}_{k, \sigma} & \equiv 
\beta_k\hat{c}_{k,\sigma}-\mathi \alpha_k\hat{c}_{k+\pi,\sigma}
\label{peiops}
\end{eqnarray}
with
\begin{eqnarray}
\alpha_{k} & = &  
\sqrt{\frac{1}{2}\left(1-\frac{\epsilon(k)}{E(k)}\right)}\; ,
\nonumber\\
\beta_{k} & = &
\sgn[\Delta(k)]
\sqrt{\frac{1}{2}\left(1+\frac{\epsilon(k)}{E(k)}\right)} \; ,
\end{eqnarray}
and $\sgn(x\neq 0)=x/|x|$ is the sign function.
For the inverse transformation and other useful relations, see~\ref{appA}.

\subsection{Coulomb interaction}
\label{subsec:CI}

The electrons are supposed to interact locally with strength~$U$.
The Hubbard inter\-action reads
\begin{equation}
\hat{U} = U \sum_{l} (\hat{n}_{l,\uparrow}-1/2)
(\hat{n}_{l,\downarrow}-1/2) \; ,
\end{equation}
where $\hat{n}_{l,\sigma}=
\hat{c}^+_{l,\sigma}\hat{c}_{l,\sigma}^{\phantom{+}}$ 
is the local density operator at site~$l$ for spin~$\sigma$. 
Screening is not very efficient in an insulator. Therefore,
we take into account the long-range Coulomb interaction of the electrons
in the form of a $1/r$ potential of effective strength~$V$,
\begin{equation}
\hat{V} = \sum_r V(r) \sum_{l} \left(\hat{n}_{l}-1\right)
\left(\hat{n}_{l+r}-1\right)
\; ,
\end{equation}
where $\hat{n}_{l}= \hat{n}_{l,\uparrow}+\hat{n}_{l,\downarrow}$ 
counts the electrons on site~$l$ and $V(r)=V/(2r)$ describes the Coulomb
potential. For the analytical calculations, the specific
form of $V(r)$ is not important.

\subsection{Extended Peierls--Hubbard model}
\label{subsec:Symm}

Altogether we investigate the one-dimensional
extended Peierls--Hubbard model,
\begin{equation}
\hat{H}=\hat{T} + \hat{U} +\hat{V} \equiv \hat{T} + \hat{W}\; .
\label{generalH}
\end{equation}
The Hamiltonian is invariant under SU(2) spin transformations
so that the total spin is a good quantum number.
Including the charge-SU(2) the Hamiltonian is SO(4) symmetric.
Moreover, it exhibits particle-hole symmetry,
i.e., $\hat{H}$ is invariant under the transformation 
\begin{equation}
{\rm PH}: \qquad \hat{c}_{l,\sigma}^+ \mapsto (-1)^{l}
\hat{c}_{l,\sigma}\quad ; \quad
\hat{c}_{l,\sigma} \mapsto (-1)^{l} \hat{c}_{l,\sigma}^+ \; .
\label{phdef}
\end{equation}
The chemical potential $\mu=0$ guarantees a half-filled band
for all temperatures~\cite{Gebhardbook}.

The Peierls dimerization ($\delta\neq 0$) and the
Coulomb interaction ($U,V\neq 0$) individually lead to an insulating
ground state at half band-filling. A field-theoretical investigation
shows~\cite{AnjaPhD,Eckern} that the lattice distortion is always a relevant
perturbation, i.e., it occurs for all values of the Coulomb 
interaction. Therefore, the Peierls insulator provides a valid starting point
for a perturbation expansion in the Coulomb interaction.

\section{Ground-state energy to second order}
\label{sec:gsener}
\subsection{Analytical results}

For the calculation of the ground-state energy $E_0$
we apply standard Rayleigh--Schr\"odinger perturbation theory to second order,
\begin{eqnarray}
E_0&=& E_0^{(0)} + E_0^{(1)}+ E_0^{(2)}\; , \nonumber \\
E_0^{(1)}&=& E_0^{U} + E_0^{V} \; , \label{masterenergie}\\
E_0^{(2)} &=& E_0^{U^2} + E_0^{V^2} + E_0^{UV} \; .  \nonumber
\end{eqnarray}
The expectation value of~$\hat{H}$ with 
the ground state of the Peierls insulator 
(filled lower Peierls band)
\begin{equation}
|{\rm FS}\rangle = \prod_{k,\sigma}\hat{a}_{k, \sigma}^+ |{\rm vacuum}\rangle
\label{defFS}
\end{equation}
contributes
\begin{eqnarray}
E_0^{(0)} &=& \langle {\rm FS}| \hat{T} |{\rm FS}\rangle =
-2\sum_{k} 2t\sqrt{\cos(k)^2+\delta^2\sin(k)^2}\; , \nonumber\\
E_0^{U} &=& \langle {\rm FS}| \hat{U} |{\rm FS}\rangle =0 \; , 
\label{ezerouiszero}\\
E_0^{V} &=& \langle {\rm FS}| \hat{V} |{\rm FS}\rangle =
-L\sum_{-L/2\leq r<L/2\atop{ r|2=1 }}
2V(r)[(A_{\delta}(r))^2+(B_{\delta}(r))^2]\; ,
\end{eqnarray}
where ($r|2=1$) denotes all odd~$r$, $A_{\delta}(r)$, $B_{\delta}(r)$
are defined in~(\ref{abdelta}), and~(\ref{onehalf}) 
is used for the derivation of~(\ref{ezerouiszero}).
For the Peierls insulator we have in the thermodynamic limit,
\begin{equation}
\lim_{L\to\infty}\frac{E_0^{(0)}}{L}= -(4t/\pi){\rm E}(1-\delta^2)
\end{equation}
where ${\rm E}(m)$ is the complete elliptic integral of the second kind.

\subsubsection{Second-order excitations.}

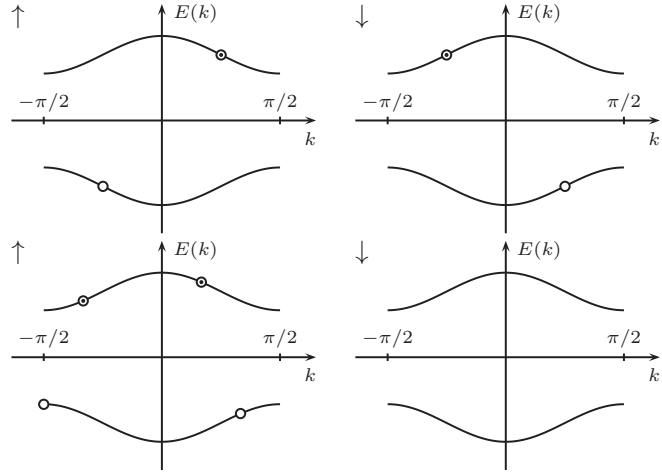
\begin{figure}[htbp]
\begin{center}
\begin{equation*}
\begin{tabular}{ll}
  \begin{pspicture}[0.382](-2.05,-1.55)(2.1,1.6)
    \anregung{$\mathbf{\uparrow}$}
    \rput(1.57080,1.75){\teilchen}
    \rput(-1.57080,-1.75){\loch}
  \end{pspicture}
  &
  \begin{pspicture}[0.382](-2.05,-1.55)(2.1,1.6)
    \anregung{$\mathbf{\downarrow}$}
    \rput(-1.57080,1.75){\teilchen}
    \rput(1.57080,-1.75){\loch}
  \end{pspicture}
  \\
  \begin{pspicture}[0.382](-2.05,-1.55)(2.1,1.6)
    \anregung{$\mathbf{\uparrow}$}
    \rput(-2.09440,1.50){\teilchen}
    \rput(1.04720,2.00){\teilchen}
    \rput(-3.14159,-1.25){\loch}
    \rput(2.09440,-1.50){\loch}
  \end{pspicture}
  &
  \begin{pspicture}[0.382](-2.05,-1.55)(2.1,1.6)
    \anregung{$\mathbf{\downarrow}$}
  \end{pspicture}
\end{tabular}
\end{equation*}
\caption{Two-particle-hole excitations $|11\rangle$ (top row)
and $|20\rangle$ (bottom row).\label{Fig:ZTLA}}
\end{center}
\end{figure}

To second order, particle-hole
excitations of the Peierls ground states must be considered.
In general, we denote excitations with $r$ ($s$)~particle-hole 
excitations in the $\uparrow$-sector ($\downarrow$-sector) by
$|rs\rangle$. Fig.~\ref{Fig:ZTLA} shows two particle-hole excitations
with antiparallel spins, i.e., $|11\rangle$, and parallel spins, i.e.,
$|20\rangle$, respectively.
These are all excitations which contribute to the ground-state energy 
to second order.

\subsubsection{Second order in the Hubbard interaction.}

For the Hubbard interaction, only the states~$|11\rangle$
contribute. As input to~(\ref{masterenergie}) 
they give~\cite{AnjaPhD}
\begin{eqnarray}
E_0^{U^2}
&=& \sum_{|11\rangle}
\frac{\left|\langle {\rm FS}|\hat{U}|11\rangle\right|^2}%
{E_0^{(0)}-E_{|11\rangle}^{(0)}}
\nonumber\\
&=&-\left(\frac{U}{L}\right)^2
\sum_{k_1,k_2,k_3,k_4 }
\frac{1}{E(k_1)+E(k_2)+E(k_3)+E(k_4)}
\nonumber\\
&&
\hphantom{-\left(\frac{U}{L}\right)^2
\sum_{k_1,k_2,k_3,k_4 } }
\Bigl\{\delta_{k_1-k_2+k_3-k_4,0}[u_1(k_1,k_2,k_3,k_4)]^2
\nonumber \\
&&\hphantom{-\left(\frac{U}{L}\right)^2 \sum_{k_1,k_2,k_3,k_4 }}
+\delta_{k_1-k_2+k_3-k_4,\pm \pi}[u_2(k_1,k_2,k_3,k_4)]^2\Bigr\}
\label{gz_u2_ztla11}
\end{eqnarray}
with $u_{1,2}(k_1,k_2,k_3,k_4)$ from~(\ref{uabbr}).

\subsubsection{Second order in the long-range interaction.}

The long-range Coulomb inter\-action also induces scatterings between particles
with the same spin. Therefore, we have three contributions to second order.
We find~\cite{AnjaPhD}
\begin{equation}
E_0^{V^2 |10\rangle} =
\sum_{|10\rangle}
\frac{\left|\langle {\rm FS}| \hat{V}|10\rangle\right|^2}%
{E_0^{(0)}-E_{|10\rangle}^{(0)}}
= \sum_k\frac{V^2 v_0(k)^2}{E(k)}\; ,
\label{gz_v2_etla}
\end{equation}
where $v_0(k)$ is given in~(\ref{v0}),
\begin{eqnarray}
E_0^{V^2 |11\rangle}
&=&
\sum_{k_1,k_2,k_3,k_4}
\frac{1}{E(k_1)+E(k_2)+E(k_3)+E(k_4)}		\nonumber\\
&&
\hphantom{\sum_{k_1,k_2,k_3,k_4}}
\Bigl\{\delta_{k_1-k_2+k_3-k_4,0}[v_1(k_1,k_2,k_3,k_4)]^2
\nonumber \\
&&\hphantom{\sum_{k_1,k_2,k_3,k_4}}
+\delta_{k_1-k_2+k_3-k_4,\pm \pi}[v_2(k_1,k_2,k_3,k_4)]^2\Bigr\}
\label{gz_v2_ztla11}
\end{eqnarray}
with $v_{1,2}(k_1,k_2,k_3,k_4)$ from~(\ref{v1}) and~(\ref{v2}),
and
\begin{eqnarray}
E_0^{V^2 |20\rangle} &=&  
-\left(\frac{V}{L}\right)^2
\sum_{  k_1<k_3,k_2<k_4 }
\frac{1}{E(k_1)+E(k_2)+E(k_3)+E(k_4)}\nonumber\\
&&
\Bigl[\delta_{k_1-k_2+k_3-k_4,0}
[v_1(k_1,k_2,k_3,k_4)-v_1(k_1,k_4,k_3,k_2)]^2\nonumber\\
&&
+ \delta_{k_1-k_2+k_3-k_4,\pm \pi}
[v_2(k_1,k_2,k_3,k_4)-v_2(k_1,k_4,k_3,k_2)]^2\Bigr]\, .
\nonumber \\
\label{gz_v2_ztla20}
\end{eqnarray}
The summation restriction in the last term
can be relaxed. Due to the symmetry of the functions 
$v_{1,2}(k_1,k_2,k_3,k_4)$
we may multiply the last contribution by a factor $1/4$
and sum over all momenta from the reduced Brillouin zone.
Due to the spin-flip symmetry we altogether have
\begin{equation}
E_0^{V^2}=2 E_0^{V^2 |10\rangle}+
E_0^{V^2 |11\rangle}+ 2 E_0^{V^2 |20\rangle}
\end{equation}
as input to~(\ref{masterenergie}).

\subsubsection{Second-order mixed interactions.}

Only the states~$|11\rangle$ contribute to order~$UV$,
\begin{eqnarray}
E_0^{UV}
&=&\sum_{|11\rangle}
\frac{\langle {\rm FS}|\hat{V}|11\rangle \langle 11|\hat{U}|{\rm FS}\rangle 
+{\rm c.c.}}{E_0^{(0)}-E_{|11\rangle}^{(0)}}
\nonumber\\
&=& \frac{2UV}{L^2}
\sum_{ k_1,k_2,k_3,k_4 }\frac{1}{E(k_1)+E(k_2)+E(k_3)+E(k_4)}		
\nonumber\\
&&
\Bigl[\delta_{k_1-k_2+k_3-k_4,0}u_1(k_1,k_2,k_3,k_4)v_1(k_1,k_2,k_3,k_4)
\nonumber\\
&&
+\delta_{k_1-k_2+k_3-k_4,\pm \pi}u_2(k_1,k_2,k_3,k_4)v_2(k_1,k_2,k_3,k_4)
\Bigr] \; .
\label{gz_uv_ztla11}
\end{eqnarray}

\subsection{Comparison with numerical results}

\subsubsection{Finite-size effects.}

\begin{figure}[htbp]
\begin{center}
\includegraphics[width=11cm]{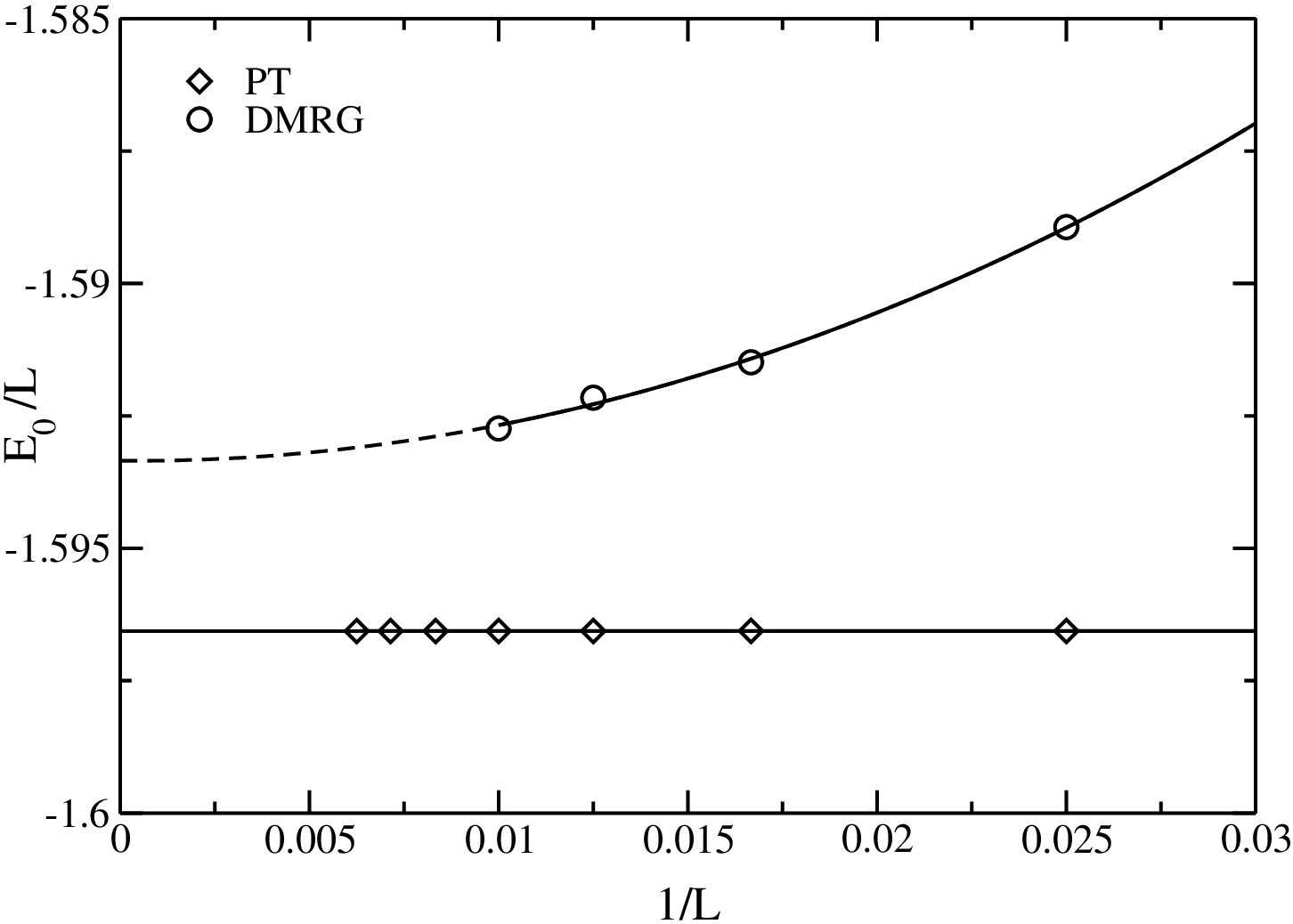}
\caption{Ground-state energy density $E_0/L$ 
as a function of inverse system size
for fixed $U=2V=2t$ and $\delta=0.2$ in the DMRG and 
perturbation theory; lines are polynomial fits. Note the resolution
of the ordinate.\label{Fig:ezerofiniteL}}
\end{center}
\end{figure}

First, we demonstrate that lattices with $L=100$ provide
results which are very close to the thermodynamic limit.
Fig.~\ref{Fig:ezerofiniteL} gives an example
for $U=2V=2t$ and $\delta=0.2$. 

The result from perturbation theory
for periodic boundary conditions shows almost no size dependence whereas 
the data from the numerical density-matrix renormalization group
displays the typical parabolic convergence in $1/L$. The comparison
shows that the results for $L=100$ are almost identical to
the result in the thermodynamic limit. In particular,
the differences between perturbation theory and the numerically exact DMRG
are small but significant. The same observation holds equally well
for other choices of (finite) model parameters $U,V,\delta$.

In the following we shall show results for $L=100$ 
and take them as representative for the thermodynamic limit.

\subsubsection{Fixed ratio $U/V$.}

\begin{figure}[htbp]
\begin{center}
\includegraphics[width=11cm]{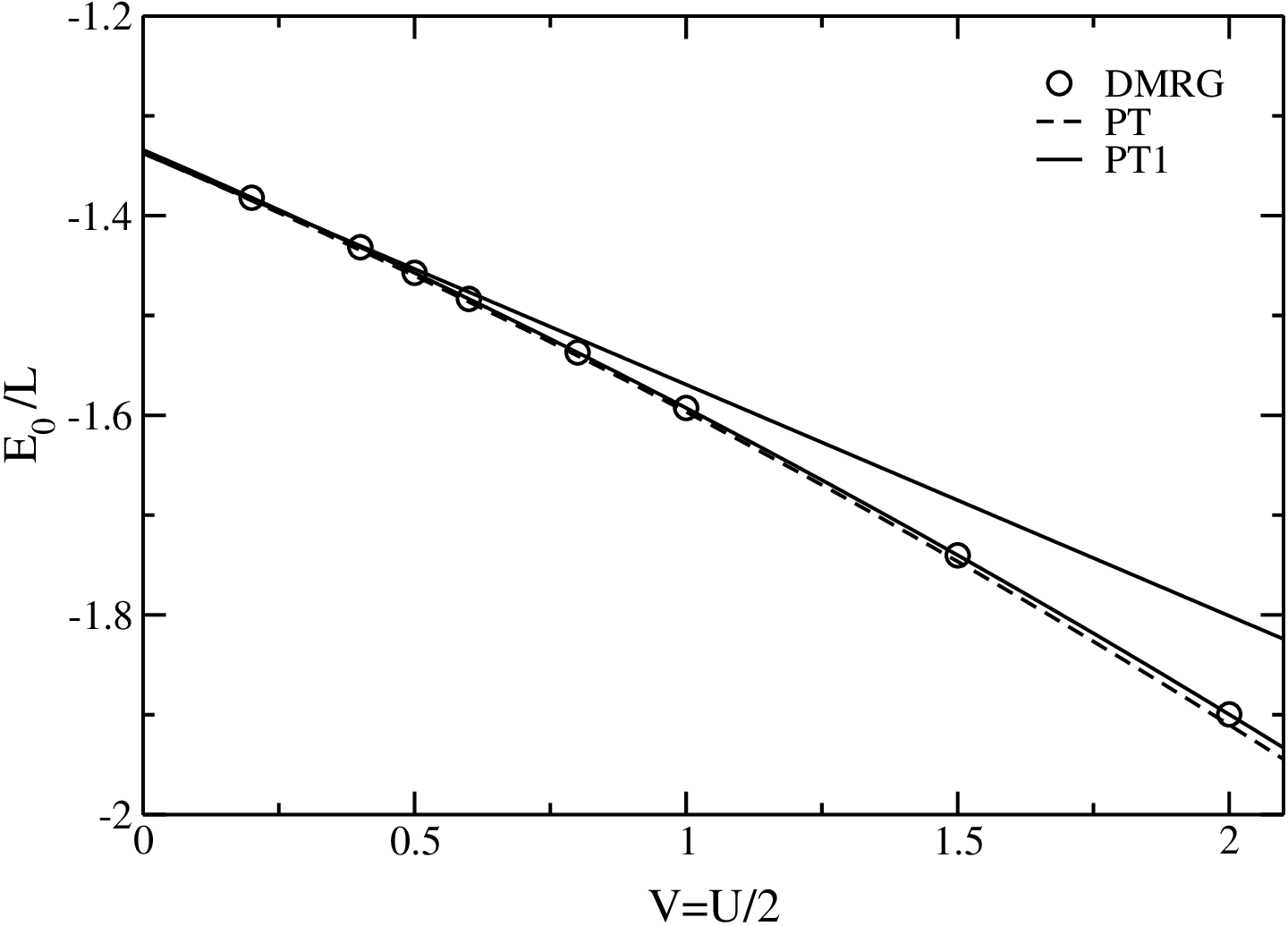}
\caption{Ground-state energy density $E_0/L$ 
as a function of~$V$ for fixed $U/V=2$, $L=100$, and $\delta=0.2$ in the DMRG 
(circles) and perturbation theory. Lines are polynomial fits.
For $U=V=0$,
$\lim_{L\to\infty}E_0^{(0)}/L\approx 1.34$.\label{Fig:gz-energy-vs-dmrg_u2v}}
\end{center}
\end{figure}

Next, we discuss the ground-state energy density as a function
of the Coulomb interaction for a fixed ratio of~$U/V$. This is shown
in Fig.~\ref{Fig:gz-energy-vs-dmrg_u2v} where we compare the DMRG data
with the results from first and second-order perturbation theory.
It is seen that the second-order correction considerably improves 
the agreement between the numerically exact DMRG data and perturbation theory.

\subsubsection{Fixed~$V$.}

\begin{figure}[htbp]
\begin{center}
\includegraphics[width=11cm]{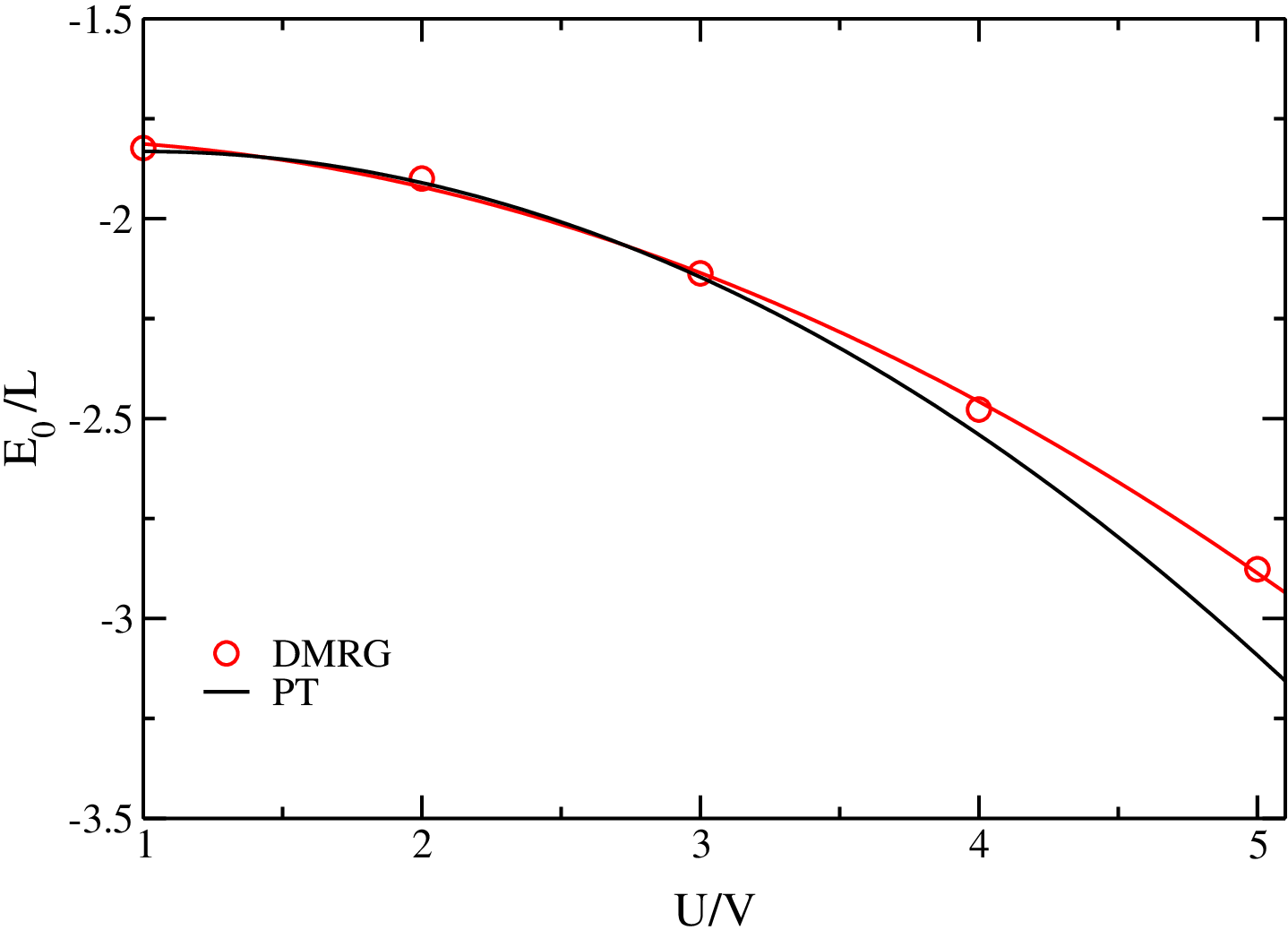}
\caption{Ground-state energy density $E_0/L$ 
as a function of~$U$ for $V=2t$, $L=100$, and $\delta=0.2$ in the DMRG 
(circles) and perturbation theory. Lines are polynomial 
fits.\label{Fig:gz-energy-vs-dmrg-v2}}
\end{center}
\end{figure}

In general, the influence of the Hubbard interaction is well
described by the  second-order term, as can be seen from 
Fig.~\ref{Fig:gz-energy-vs-dmrg-v2}. It shows that perturbation theory
works very well even when $U$~is as large as the band width $W=4t$.
The agreement is even better for smaller~$V$.

\subsubsection{Conclusion.}

The above comparison has been made for a sizable value of the Peierls
dimerization, $\delta=0.2$, so that the bare Peierls gap,
$\Delta^{\rm P}=4t\delta=0.8t$, is substantial.
Naturally, the comparison between the DMRG and second-order perturbation theory
becomes less favorable when we decrease~$\delta$ to realistic values,
$\delta< 0.1$; see section~\ref{sec:discussion}. 
Nevertheless, even for small values of~$\delta$,
the formulae in this section give
a very good estimate for the ground-state energy density for a large
range of Coulomb interactions, $U\approx 2V \leq W$.

\section{Single-particle gap to second order}
\label{sec:singleP}

In order to generate current-carrying excitations in a Peierls--Mott 
insula\-tor,
the gap for single-particle excitations must be overcome.
It is defined by 
\begin{equation}
\Delta_1 = [E_0(L+1)-E_0(L)]-[E_0(L)-E_0(L-1)]\; ,
\end{equation}
where $E_0(N)$ is the ground-state energy of the system with
$N$~electrons. Due to particle-hole symmetry we have
\begin{equation}
\Delta_1 = 2[E_0(L+1)-E_0(L)]\; .
\end{equation}

\subsection{Analytical results}

For the perturbative calculation of the single-particle gap~$\Delta_1$
we apply standard Rayleigh--Schr\"odinger perturbation theory again and write
\begin{equation}
\Delta_1= \Delta_1^{(0)} + \Delta_1^{U} + \Delta_1^{V} 
+ \Delta_1^{U^2} + \Delta_1^{V^2} + \Delta_1^{UV} 
\; . \label{masterdelta}
\end{equation}
For the Peierls insulator the first two contributions are readily calculated
because the ground state with $N=L+1$ particles is
\begin{equation}
|p\rangle \equiv 
\hat{b}_{p,\uparrow}^+|{\rm FS}\rangle \quad , \quad p=-\pi/2 \; ,
\end{equation}
and, thus,
\begin{equation}
\Delta_1^{(0)} = 2 \left[\langle p |\hat{T} | p \rangle - \langle {\rm FS} | \hat{T} |
{\rm FS} \rangle\right] = 2E(p) 
=4t\delta \equiv \Delta^{\rm P}
\end{equation}
is the Peierls gap for all system sizes. Due to the spin symmetry we were
allowed to use $\sigma=\uparrow$.
Moreover, the first-order contributions read
\begin{eqnarray}
\Delta_1^{U} &=& 
2 \left[\langle p | \hat{U} | p \rangle 
- \langle {\rm FS}| \hat{U} |{\rm FS}\rangle\right]
 =0 \; , 
\label{etl_u}\\
\Delta_1^{V} &=& 
2 \left[\langle p | \hat{V} | p \rangle 
- \langle {\rm FS}| \hat{V} |{\rm FS}\rangle\right]
\nonumber \\
&=& 2\sum_{-L/2< r<L/2\atop{r|2=1}} 4tV(r)
\frac{\cos(p)\cos(p r)A_{\delta}(r)+
\delta\sin(p)\sin(p r)B_{\delta}(r)}{E(p)}\nonumber\\
&=&
\sum_{-L/2< r < L/2\atop{r|2=1}}4V(r)(-1)^{[(r-1)/2]}  B_{\delta}(r)\;,
\label{etl_v}
\end{eqnarray}
where $p=-\pi/2$ was used in the last step.

\subsubsection{Second order in the Hubbard interaction.}

In the intermediate states an additional $\uparrow$ electron
and $r$ ($s$) electron-holes excitations in the $\uparrow$-sector
($\downarrow$-sector) are possible, denoted as $|k;rs\rangle$.
There are two contributions to second order in the Hubbard interaction,
\begin{eqnarray}
\Delta_1^{U^2 |01\rangle}
&=& 2\sum_{k_1,k_2,k_3}
\frac{|\langle p|\hat{U}
\hat{b}_{k_1,\downarrow}^{+}\hat{a}_{k_2,\downarrow}|k_3\rangle|^2}
{E(p)-(E(k_1)+E(k_2)+E(k_3))} \nonumber \\
&=&  \left(\frac{U}{L}\right)^2
\sum_{k_1,k_2,k_3}
\frac{2\left(1-\delta_{p,k_1}\right)}{E(p)-(E(k_1)+E(k_2)+E(k_3))}\nonumber\\
&&\hphantom{\left(\frac{U}{L}\right)^2 \sum_{k_1,k_2,k_3}}
\Bigl[\delta_{k_1-k_2+k_3-p,0}[u_2(k_1,k_2,k_3,p)]^2 \nonumber \\
&&\hphantom{\left(\frac{U}{L}\right)^2 \sum_{k_1,k_2,k_3}}
+\delta_{k_1-k_2+k_3-p,\pm \pi}[u_1(k_1,k_2,k_3,p)]^2\Bigr]\; ,
\label{etl_u2_etla}
\end{eqnarray}
and 
\begin{eqnarray}
\frac{\Delta_1^{U^2 |11\rangle}}{2}
&=& \sum_{ k_1,k_2,k_3,k_4 }
\frac{|\langle p | \hat{U}
\hat{b}_{k_1,\uparrow}^{+}\hat{a}_{k_2,\uparrow}
\hat{b}_{k_3,\downarrow}^{+}\hat{a}_{k_4,\downarrow}|p\rangle|^2}
{-\left(E(k_1)+E(k_2)+E(k_3)+E(k_4)\right)}
-E_0^{U^2|11\rangle}\nonumber \\
&=&\left(\frac{U}{L}\right)^2 \sum_{ k_1,k_2,k_3,k_4 }
\frac{\delta_{p,k_1}}{E(k_1)+E(k_2)+E(k_3)+E(k_4)}\nonumber\\
&&\hphantom{\left(\frac{U}{L}\right)^2 \sum_{ k_1,k_2,k_3,k_4 }}
\Bigl[\delta_{k_1-k_2+k_3-k_4,0}[u_1(k_1,k_2,k_3,k_4)]^2 \label{etl_u2_ztla11}
\\
&&\hphantom{\left(\frac{U}{L}\right)^2 \sum_{ k_1,k_2,k_3,k_4 }}
+\delta_{k_1-k_2+k_3-k_4,\pm \pi}[u_2(k_1,k_2,k_3,k_4)]^2\Bigr]\; .
\nonumber
\end{eqnarray}
We thus find from these two equations
\begin{equation}
\Delta_1^{U^2}= 
\Delta_1^{U^2 |01\rangle} +
\Delta_1^{U^2 |11\rangle} 
\end{equation}
as input to~(\ref{masterdelta}).

\subsubsection{Second order in the long-range interaction.}

To second order in the long-range interaction, the states
$|k;10\rangle$ and $|k;01\rangle$ contribute. Note that
the fermionic nature of the operators must be taken into account
properly, e.g.,
$\hat{b}_{k,\uparrow}^+|p\rangle \equiv - \hat{b}_{p,\uparrow}^+|k\rangle$.
This gives the contributions
\begin{eqnarray}
\Delta_1^{V^2|10\rangle}&=&  
\sum_{k_2,k_1>k_3}
\frac{2|\langle p|
\hat{V}\hat{b}_{k_1,\uparrow}^{+}\hat{a}_{k_2,\uparrow}
|k_3\rangle|^2}{E(p)-(E(k_1)+E(k_2)+E(k_3))}
-2E_0^{V^2|10\rangle} \nonumber \\
&=& \left(\frac{V}{L}\right)^2\sum_{k_2,k_1>k_3}
\frac{2}{E(p)-(E(k_1)+E(k_2)+E(k_3))}
\nonumber\\
&&\biggl[\delta_{k_1-k_2+k_3-p,0}
\Bigl\{\delta_{p,k_3}\delta_{k_1,k_2}L v_0(k_1)
-\delta_{p,k_1}\delta_{k_2,k_3}L v_0(k_3) 
\nonumber \\
&& \hphantom{\biggl[\delta_{k_1-k_2+k_3-p,0}\Bigl\{}
+ v_2(k_3,p,k_1,k_2)-v_2(k_1,p,k_3,k_2)\Bigr\}^2 
\nonumber\\
&&\hphantom{\biggl[}
+\delta_{k_1-k_2+k_3-p,\pm \pi}
\Bigl\{ v_1(k_3,p,k_1,k_2)-v_1(k_1,p,k_3,k_2)\Bigr\}^2 \biggr]
\nonumber \\
&& -2E_0^{V^2|10\rangle} \; ,
\label{etl_v2A_etla}
\end{eqnarray}
and
\begin{eqnarray}
\frac{\Delta_1^{V^2|01\rangle}}{2}&=&  
\sum_{k_1,k_2,k_3}\frac{|\langle p|
\hat{V}\hat{b}_{k_1,\downarrow}^{+}\hat{a}_{k_2,\downarrow}|k_3\rangle|^2}%
{E(p)-(E(k_1)+E(k_2)+E(k_3))}-E_0^{V^2|01\rangle} \nonumber \\
&=& \left(\frac{V}{L}\right)^2
\sum_{k_1,k_2,k_3}\frac{1}{E(p)-(E(k_1)+E(k_2)+E(k_3))}
\nonumber\\
&&\Bigl[\delta_{k_1-k_2+k_3-p,0}\Bigl\{\delta_{p,k_3}\delta_{k_1,k_2}Lv_0(k_1)+
 v_2(k_3,p,k_1,k_2)\Bigr\}^2 
\nonumber\\
&&+\delta_{k_1-k_2+k_3-p,\pm \pi}\Bigl\{ v_1(k_3,p,k_1,k_2)\Bigr\}^2 \Bigr]
\nonumber \\
&& -E_0^{V^2|01\rangle}\; .
\label{etl_v2B_etla}
\end{eqnarray}
Moreover, the states with two particle-hole excitations, $|p;11\rangle$,
contribute.
We find
\begin{eqnarray}
\frac{\Delta_1^{V^2|11\rangle}}{2}&=&
\sum_{k_1,k_2,k_3,k_4} \frac{|\langle p|
\hat{V}\hat{b}_{k_1, \uparrow}^{+}\hat{a}_{k_2, \uparrow}
\hat{b}_{k_3, \downarrow}^{+}\hat{a}_{k_4, \downarrow}
|p\rangle|^2}{-\left(E(k_1)+E(k_2)+E(k_3)+E(k_4)\right)} 
-E_0^{V^2|11\rangle} \nonumber\\
&=&\left(\frac{V}{L}\right)^2\sum_{ k_1,k_2,k_3,k_4 }
\frac{\delta_{k_1,p}}{E(k_1)+E(k_2)+E(k_3)+E(k_4)}\nonumber\\
&&\hphantom{\left(\frac{V}{L}\right)^2\sum_{ k_1,k_2,k_3,k_4 }}
\Bigl[\delta_{k_1-k_2+k_3-p,0}[v_1(k_1,k_2,k_3,k_4)]^2  \label{etl_v2_ztla11}\\
&& \hphantom{\left(\frac{V}{L}\right)^2\sum_{ k_1,k_2,k_3,k_4 }}
+\delta_{k_1-k_2+k_3-p,\pm \pi}[v_2(k_1,k_2,k_3,k_4)]^2\Bigr] \; ,
\nonumber 
\end{eqnarray}
and
\begin{eqnarray}
\frac{\Delta_1^{V^2|20\rangle}}{2}&=&
\sum_{{k_1<k_3}\atop{k_2<k_4}} 
\frac{|\langle p|\hat{V}\hat{b}_{k_1,\uparrow}^{+}\hat{a}_{k_2,\uparrow}
\hat{b}_{k_3,\uparrow}^{+}\hat{a}_{k_4,\uparrow}|p\rangle|^2}%
{-\left(E(k_1)+E(k_2)+E(k_3)+E(k_4)\right)}-E_0^{V^2|20\rangle}\nonumber \\
&=&\left(\frac{V}{L}\right)^2\sum_{ k_1<k_3,k_2<k_4}
\frac{\delta_{k_1,p}+\delta_{k_3,p}}{E(k_1)+E(k_2)+E(k_3)+E(k_4)}
\nonumber\\
&&
\Bigl[\delta_{k_1-k_2+k_3-p,0}\left[v_1(k_1,k_2,k_3,k_4)-v_1(k_1,k_4,k_3,k_2) 
\right]^2 \nonumber \\
&&
+\delta_{k_1-k_2+k_3-p,\pm \pi}[v_2(k_1,k_2,k_3,k_4)-v_2(k_1,k_4,k_3,k_2)]^2
\Bigr]
\nonumber \\
\label{etl_v2_ztla20}
\end{eqnarray}
with $\Delta_1^{V^2|02\rangle}=\Delta_1^{V^2|20\rangle}$.
Altogether, from~(\ref{etl_v2A_etla}), (\ref{etl_v2B_etla}),
(\ref{etl_v2_ztla11}), and~(\ref{etl_v2_ztla20}) we find
\begin{equation}
\Delta_1^{V^2} = 
\Delta_1^{V^2|10\rangle} +
\Delta_1^{V^2|01\rangle} +
\Delta_1^{V^2|11\rangle} + 
2\Delta_1^{V^2|20\rangle}
\end{equation}
as input to~(\ref{masterdelta}).

\subsubsection{Second-order mixed interactions.}

As for the Hubbard interaction we have two contributions for the $UV$~mixed
interactions. From the inter\-mediate states~$|k;01\rangle$ we obtain
\begin{eqnarray}
\Delta_1^{UV|01\rangle}&=& \sum_{k_1,k_2,k_3}
\frac{2}{E(p)-(E(k_1)+E(k_2)+E(k_3))} 
\nonumber \\
&&\Bigl[\langle p|\hat{U}\hat{b}_{k_1,\downarrow}^{+}\hat{a}_{k_2,\downarrow}
|k_3\rangle
\langle p|\hat{V}\hat{b}_{k_1,\,\downarrow}^{+}\hat{a}_{k_2,\,\downarrow}
|k_3\rangle^{*} +{\rm c.c.}\Bigr]
\nonumber\\
&=& \left(\frac{4UV}{L^2}\right)
\sum_{ k_1,k_2,k_3 }\frac{1}{E(p)-(E(k_1)+E(k_2)+E(k_3))} 
\nonumber\\
&&
\Bigl[\delta_{k_1-k_2+k_3-p,0}v_2(k_3,p,k_1,k_2)u_2(k_1,k_2,k_3,p) 
\nonumber\\
&&+\delta_{k_1-k_2+k_3-p,\pm \pi}v_1(k_3,p,k_1,k_2)u_1(k_1,k_2,k_3,p)\Bigr]\;,
\label{etl_uv_etla}
\end{eqnarray}
and the states $|p;11\rangle$ give
\begin{eqnarray}
\Delta_1^{UV|11\rangle}&=&
\sum_{k_1,k_2,k_3,k_4}
\frac{(-2)}{ E(k_1)+E(k_2)+E(k_3)+E(k_4) } 
\nonumber \\
&&\hphantom{\sum}
\Bigl[\langle p|\hat{U}\hat{b}_{k_1,\uparrow}^{+}\hat{a}_{k_2,\uparrow}
\hat{b}_{k_3,\downarrow}^{+}\hat{a}_{k_4,\downarrow}|p\rangle
\langle p|\hat{V}\hat{b}_{k_1,\uparrow}^{+}\hat{a}_{k_2,\uparrow}
\hat{b}_{k_3,\downarrow}^{+}\hat{a}_{k_4,\downarrow}|p\rangle^{*} \nonumber \\
&&\hphantom{\sum\Bigl[}
+{\rm c.c.}\Bigr] -E_0^{UV|11\rangle}\nonumber \\
&=&\left(\frac{4UV}{L^2}\right)
\sum_{k_1,k_2,k_3,k_4}
\frac{ \delta_{k_1,p} }{E(k_1)+E(k_2)+E(k_3)+E(k_4)}
\nonumber\\
&& 
\Bigl[\delta_{k _1-k_2+k_3-p,0}u_1(k_1,k_2,k_3,k_4)v_1(k_1,k_2,k_3,k_4) 
\nonumber\\
&&+\delta_{k_1-k_2+k_3-p,\pm \pi}u_2(k_1,k_2,k_3,k_4)v_2(k_1,k_2,k_3,k_4)
\Bigr]\; .
\label{etl_uv_ztla11}
\end{eqnarray}
Altogether we thus find from these two equations
\begin{equation}
\Delta_1^{UV}= 
\Delta_1^{UV|01\rangle} +
\Delta_1^{UV|11\rangle}
\end{equation}
as input to~(\ref{masterdelta}).

\subsection{Comparison with numerical results}

\subsubsection{Finite-size effects.}

\begin{figure}[htbp]
\begin{center}
\includegraphics[width=12cm]{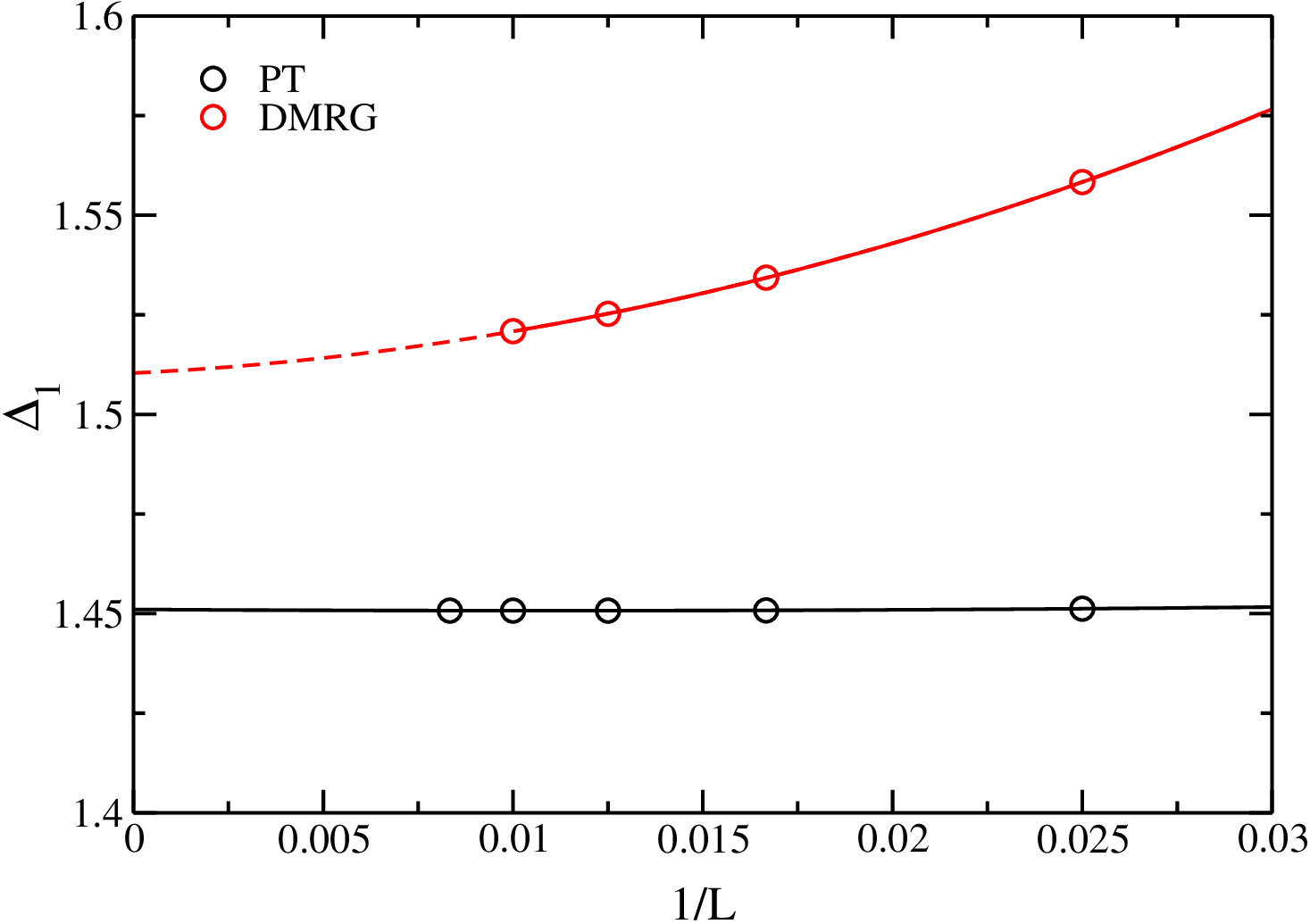}
\caption{Single-particle gap $\Delta_1$ as a function of the inverse system
size for fixed $U=2V=2t$ and $\delta=0.2$ in the DMRG and perturbation theory;
lines are polynomial fits.\label{Fig:etl-energy-vs-l}}
\end{center}
\end{figure}

As for the ground-state energy we first investigate the size dependence.
Fig.~\ref{Fig:etl-energy-vs-l} shows that the analytical results for
the single-particle gap are almost independent of the system size, 
as expected. The DMRG data show the typical quadratic convergence
as a function of $1/L$. For our assessment of the quality 
of the perturbative approach we note that the DMRG data for $L=100$
represent the value in the thermodynamic limit to an accuracy of about
one percent. Therefore, the differences between perturbation theory 
and the numerically exact DMRG which we will discuss next
are significant.

\subsubsection{Fixed ratio $U/V$.}

\begin{figure}[htb]
\begin{center}
\includegraphics[width=11cm]{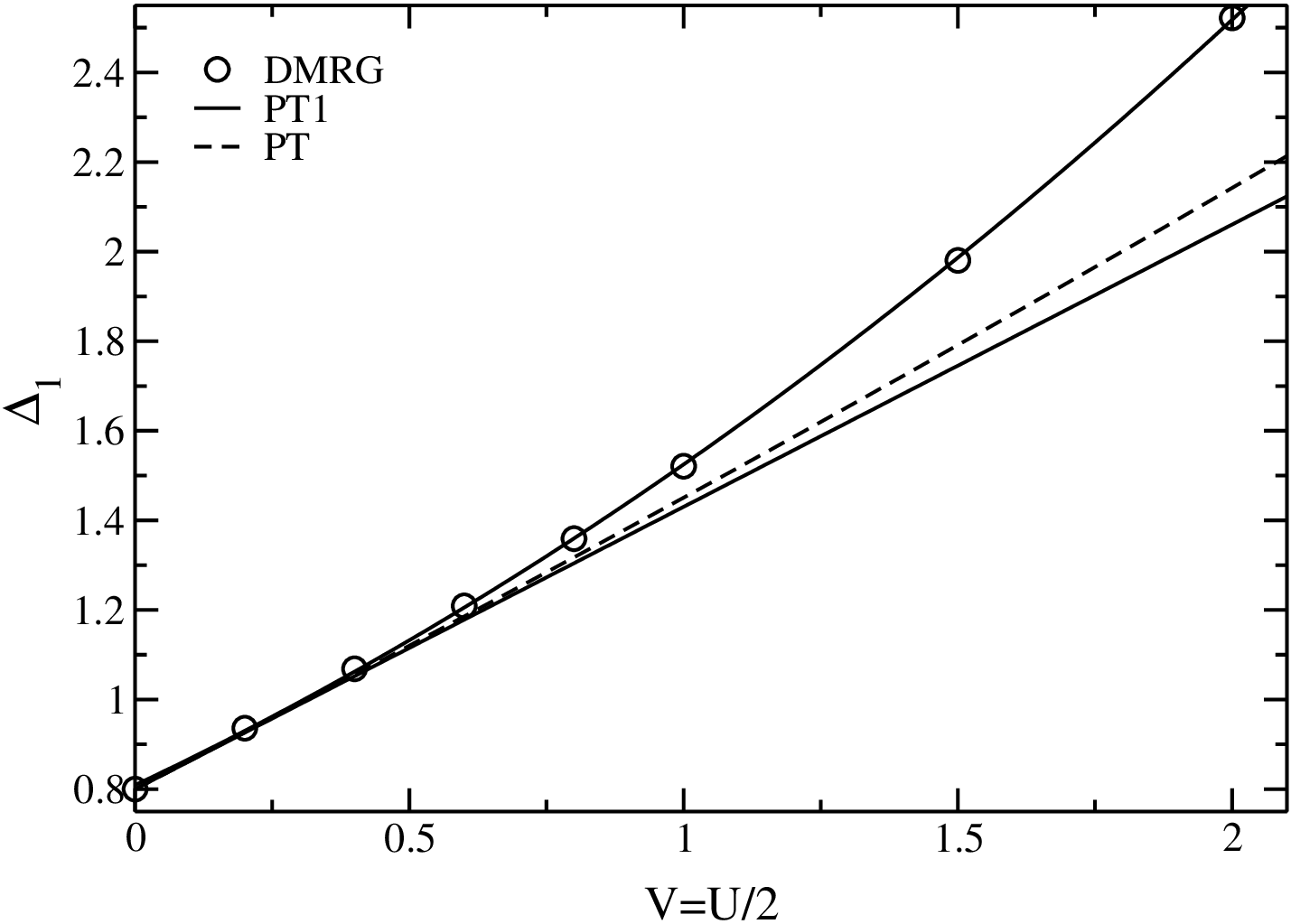}
\caption{Single-particle gap $\Delta_1$ as a function 
of~$V$ for fixed $U/V=2$, $L=100$, and $\delta=0.2$ in the DMRG (circles) and 
perturbation theory. Lower curve: perturbation theory to first order (PT1), 
upper curve: perturbation theory to second order (PT).
Lines are polynomial fits.\label{Fig:etl-energy-vs-dmrg}}
\end{center}
\end{figure}

Fig.~\ref{Fig:etl-energy-vs-dmrg} compares the single-particle gap
as a function of~$V$ for fixed ratio $U/V=2$
in the DMRG and perturbation theory to first and second order.
The agreement is worse than for the ground-state energy.
Nevertheless, the deviation between the numerically exact DMRG results
and perturbation theory at $U=3t$ and $V=1.5t$ is only about
ten percent despite the fact that the gap has increased from its
noninteracting value~$\Delta^{\rm P}=0.8t$ by a factor of~2.5 to
$\Delta_1^{\rm DMRG}(U=3t,V=1.5t,\delta=0.2)\approx 2.0t$.

\begin{figure}[htb]
\begin{center}
\includegraphics[width=11cm]{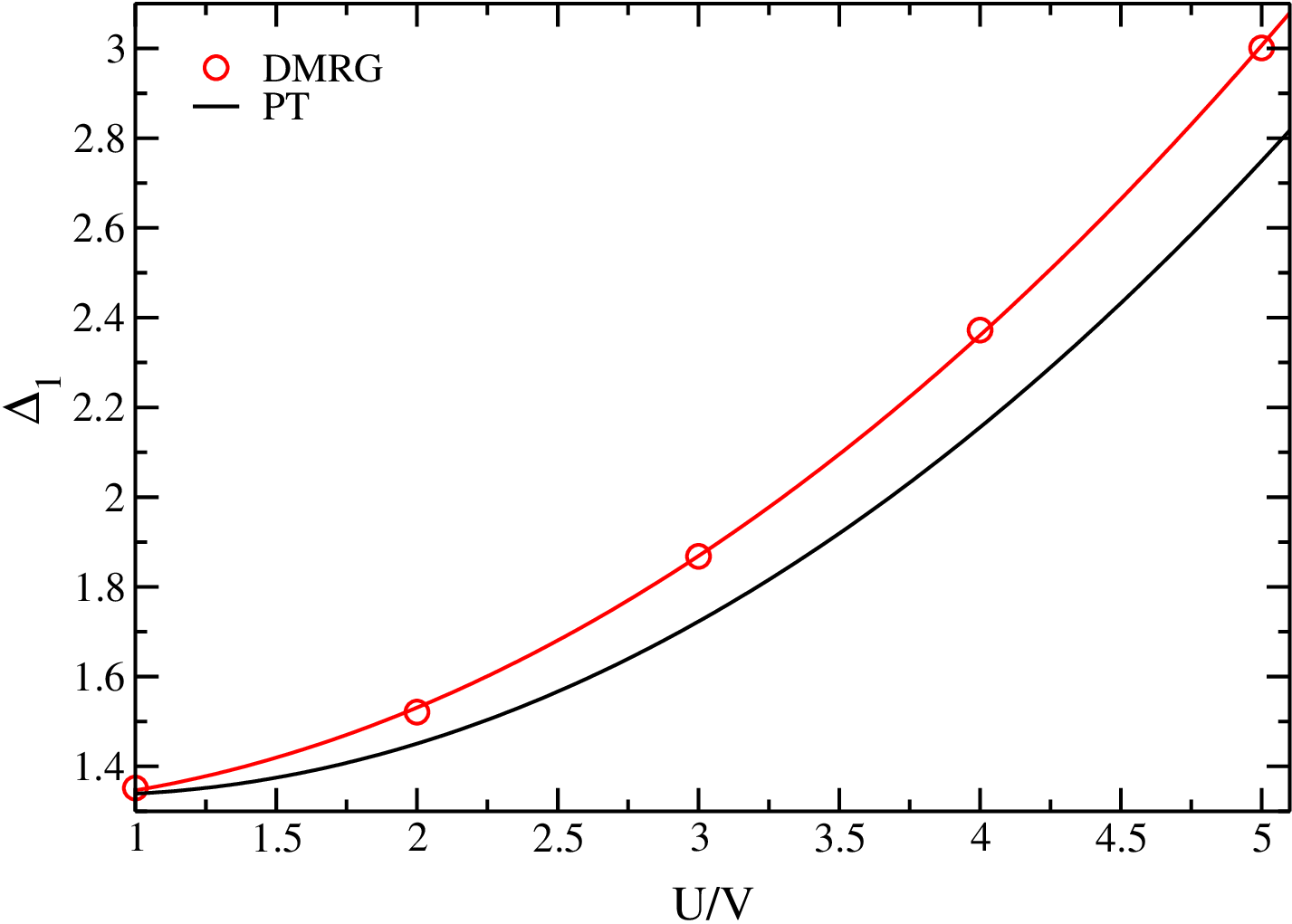}
\caption{Single-particle gap $\Delta_1$ as a function of $U/V$ for $V=t$, 
$L=100$, and  $\delta=0.2$ in the DMRG (circles) and perturbation 
theory; lines are polynomial fits.\label{Fig:etl-energy-vs-dmrg-v1}}
\end{center}
\end{figure}

The figure also shows that the second-order contribution only slightly improves
the first-order result. The surprisingly good first-order result is partially
due to the error compensation in the calculation of the ground-state
energies for $N=L$ and $N=L+1$ particles.  

\subsubsection{Fixed~$V$.}

As in the case of the ground-state energy, the discrepancies between
the DMRG single-particle gap and the result from second-order
perturbation theory
are mainly caused by the long-range part of the Coulomb interaction.
This is shown in Fig.~\ref{Fig:etl-energy-vs-dmrg-v1}. 
The results from second-order perturbation theory
closely follow the DMRG data for all $U/t$ at fixed~$V=t$.

At $U=3t$ and $V=t$ the difference between
$\Delta_1^{\rm DMRG}(U=3t,V=t,\delta=0.2)=1.8680t$ and
$\Delta_1^{\rm PT}(U=3t,V=t,\delta=0.2)=1.7229t$ is only eight percent.
Here, first-order perturbation theory gives 
$\Delta_1^{\rm PT1}(U=3t,V=t,\delta=0.2)=1.430t$ which is off by
23~percent. This shows that the inclusion of the second-order terms
improves the agreement considerably for all~$U/t$ as long as~$V/t$
is not too large.

\subsubsection{Conclusion.}

Second-order perturbation theory provides a reliable estimate
for the single-particle gap in the regime where it has almost
tripled from its bare Peierls value, i.e., in the region
($U\leq 4t$, $V\leq 2t$) for $\delta=0.2t$ ($\Delta^{\rm P}=0.8t$).
This is rather important because it indicates that only 
a minor part of the single-particle gap
in polymers is caused by the Peierls effect. For poly-acetylene
where $t\approx 2\, {\rm eV}$ and $\Delta_1\approx 1.6\, {\rm eV}$
one would obtain $\delta=0.2$ in the absence of Coulomb correlations.
When we assume a correlation enhancement for the gap
by a factor of about three, we would estimate $\delta\approx 0.07$,
in agreement with other estimates in the literature.

A reliable estimate for the single-particle gap is important for the
cal\-culation of binding energies of bound particle-hole excitations
(excitons). They are the subject of the next section.

\section{Perturbative approaches to optical excitations}
\label{sec:ptapproachestoexc}

There are various possibilities to set up second-order perturbation
theory for optical excitations. In this work we discuss Wannier, two-step,
and down-folding perturbation theories separately before we calculate
matrix elements.
Starting point is the subspace of optical excitations in the Peierls
model,
\begin{eqnarray}
{\rm singlet}: \quad| s\rangle &=&\sum_k w_s(k)  |k\rangle_s\;,
\nonumber\\
{\rm triplet}: \quad| t\rangle &=&\sum_k w_t(k) | k \rangle_t\; ,
\label{exz_wf}
\end{eqnarray}
with the normalization $\sum_{k} [w_{(s,t)}(k)]^2=1$ and 
\begin{equation}
|k\rangle_{(s,t)}\equiv \sqrt{\frac{1}{2}}\left(\hat{b}_{k, \uparrow}^+  
\hat{a}_{k, \uparrow} 
\pm\hat{b}_{k, \downarrow}^+\hat{a}_{k, \downarrow}\right)| {\rm FS}\rangle\; .
\label{ks_kt}
\end{equation}
Here, we introduce linear combinations of single particle-hole excitations
$|10\rangle$ and $|01\rangle$
from the outset because we are mainly interested in bound states
of particle and hole.

\subsection{Wannier perturbation theory}

Let $\hat{P}$ and $\hat{Q}$ be projectors onto separate subspaces
of the Hilbert space, $\hat{P}+\hat{Q}=\hat{1}$. In the case of excitons
we let $\hat{P}$ project onto the subspace of a single
particle-hole excitation, $\hat{P}\equiv \hat{P}_1$. 
Then, we may write for 
$\hat{H}=\hat{T}+\hat{W}$
\begin{eqnarray}
\hat{H}&=&\left[\hat{T}+\hat{P}\hat{W}\hat{P}\right]+
\left[\hat{Q}\hat{W}\hat{P}+\hat{P}\hat{W}\hat{Q}
+\hat{Q}\hat{W}\hat{Q}\right]\nonumber\\
&\equiv&  \hat{H}_0+\hat{H}_{\perp}\;,
\end{eqnarray}
where $\hat{H}_{0}$ describes the unperturbed system and 
$\hat{H}_{\perp}$ represents the perturbation.

\subsubsection{Wannier perturbation theory to first order.}

In the usual Wannier theory, one diagonalizes $\hat{H}_0$ in the subspace
of a single particle-hole excitation,
\begin{equation}
\hat{H}_0 | (s,t)\rangle_n = \epsilon_n^{(s,t)}  | (s,t)\rangle_n\; ,
\end{equation}
where $\epsilon_n^{(s,t)}$ is the Wannier spectrum in the spin singlet
and triplet sector. Typically, the states with lowest energy correspond to
bound states (excitons)~\cite{HaugKoch}.

\subsubsection{Wannier perturbation theory to second order.}

Given the Wannier spec\-trum and states to first order, we can calculate
the Coulomb corrections to the Wannier eigenstates via perturbation
theory to second order in~$\hat{H}_{\perp}$
\begin{equation}
E_n = \epsilon_n + \sum_{|m\rangle}
\frac{\langle x_n|\hat{H}_{\perp}|m\rangle
\langle m|\hat{H}_{\perp}|x_n\rangle}{\epsilon_n-E_m^{(0)}}
\label{2O}
\end{equation}
in the singlet and triplet sector, $x=(s,t)$.
Here, $E_m^{(0)}=\langle m|\hat{H}_0|m\rangle$, and $|m\rangle$ can be chosen
as an eigenstate of the Peierls Hamiltonian~$\hat{T}$ because
corrections are of higher order in the Coulomb interaction.
The states $|m\rangle$ contain up to three particle-hole 
excitations.
The corresponding matrix elements in~(\ref{2O})
are calculated in the next section.

\subsection{Two-step perturbation theory}

The Wannier theory for excitons probes how a bound 
excitation of bare particles and holes 
is perturbed by the Coulomb interaction. In two-step
perturbation theory the two steps are reversed: excitons are formed
from dressed particles and holes. This two-step approach is very similar
in spirit to the GW-BSE approach where the Bethe--Salpeter Equation is
solved for quasi-particles which are calculated in the 
GW-approximation~\cite{GWBSE}.

To set up the two-step perturbation theory in general, we
split the Coulomb interaction into two parts,
$\hat{W}=\hat{V}_A+\hat{V}_B$. 
Let $\hat{P}_r$ 
be the projection operators onto subspaces
with $r$~particle-hole pairs and $\hat{Q}_r=\hat{1}-\hat{P}_r$
its complement. We identify
\begin{eqnarray}
\hat{V}_A &=& \hat{P}_r\hat{W}\hat{Q}_r+
\hat{Q}_r\hat{W}\hat{P}_r+\hat{Q}_r\hat{W}\hat{Q}_r\;, \\
\hat{V}_B &=&\hat{P}_r\hat{W}\hat{P}_r\; . 
\end{eqnarray}
In the first step of the perturbation theory we calculate the
change of an eigenstate~$|n^0\rangle$ of~$\hat{T}$ due to
the influence of the perturbation~$\hat{V}_A$.
In Rayleigh--Schr\"odinger perturbation theory for 
$\hat{H}_A=\hat{T}+\hat{V}_A$
to first order we find
\begin{equation}
|n\rangle= Z_n\left\{
|n^0\rangle
+\sum_{|m\rangle}\frac{|m\rangle\langle m|\hat{V}_A|n^0\rangle}%
{E_{n^0}^{(0)}-E_m^{(0)}}\right\}\; .
\end{equation}
To second order the normalization factor $Z_n$ reads
\begin{equation}
|Z_n|^2 = 1- \sum_{|m\rangle}
\frac{\left|\langle n^0|\hat{V}_A|m\rangle\right|^2}%
{\left(E_{n^0}^{(0)}-E_m^{(0)}\right)^2}\;.
\end{equation}
The states~$|n\rangle$ are so-called `dressed states' 
because they contain particle-hole
excitations which are renormalized by the interaction~$\hat{V}_A$.

In the second step we apply (almost degenerate) perturbation theory
for the perturbation~$\hat{V}_B$. Therefore, we diagonalize 
$\hat{H}=\hat{H}_A+\hat{V}_B$ in the basis of the states $|n\rangle$,
i.e., we calculate the eigenvalues of the Hamilton matrix
\begin{equation}
H(n_1,n_2)= \langle n_1|\hat{H}|n_2\rangle
\label{twostep}
\end{equation}
in the respective subspaces with $r$~particle-hole excitations.

As an example we consider the case~$r=0$, i.e., we calculate
the second-order ground-state energy in two-step perturbation theory.
Since $|0\rangle$ is not degenerate, we have
\begin{eqnarray}
|0^0\rangle 	&=& |{\rm FS}\rangle\; ,\\
|0\rangle	&=& Z_0(|{\rm FS}\rangle+|X\rangle)\;.
\end{eqnarray}
The Hamilton matrix is a scalar,
\begin{eqnarray}
E_0 &=& \langle 0|\hat{H}|0\rangle \nonumber\\
&=& \left(|Z_0|^2-1\right)\langle{\rm FS}|\hat{T}|{\rm FS}\rangle 
+\langle{\rm FS}|\hat{T}+\hat{W}|{\rm FS}\rangle \nonumber \\
&& +\langle X|\hat{W}|{\rm FS}\rangle 
+\langle{\rm FS}|\hat{W}|X\rangle 
+\langle X|\hat{T}| X \rangle \nonumber\\
&=& \langle{\rm FS}|\hat{T}+\hat{W}|{\rm FS}\rangle
+\sum_{|m\rangle\neq |{\rm FS}\rangle}
\frac{\left|\langle {\rm FS}|\hat{W}|m\rangle\right|^2}
{E_0^{(0)}-E_m^{(0)}}\; .
\end{eqnarray}
This expression is identical to Rayleigh--Schr\"odinger perturbation theory
to second order.

\subsection{Down-folding perturbation theory}

\subsubsection{Equivalence to Brillouin--Wigner perturbation theory.}

The so-called `down-folding' approach is also based on a projection
to relevant degrees of freedom. Consider the stationary
Schr\"odinger equation at energy~$E$
\begin{equation}
\hat H |\Psi \rangle= 
\left(\hat T+\hat W\right) |\Psi \rangle= E |\Psi \rangle\; .
\label{schroedinger}
\end{equation}
For any two projectors $\hat{P}$ and $\hat{Q}$ with $\hat{P}+\hat{Q}=\hat{1}$
we define the orthogonal states
\begin{equation}
|\varphi \rangle=\hat{P}|\Psi \rangle \quad , \quad
|\chi \rangle=\hat{Q}|\Psi \rangle  
\end{equation}
which obey the coupled Schr\"odinger equations
\begin{eqnarray}
E|\varphi \rangle&=& \hat{P} \hat{H} |\varphi \rangle
+ \hat{P} \hat{H}|\chi \rangle\; ,
\nonumber\\
E|\chi \rangle&=& \hat{Q} \hat{H} |\varphi \rangle
+ \hat{Q} \hat{H}|\chi \rangle\; .
\end{eqnarray}
The second equation is formally inverted with the help of the resolvent
operator $(z-\hat{H})^{-1}$ to give a single equation
\begin{equation} 
\hat{P} \hat{H} \hat{P}|\varphi \rangle 
+ \hat{P}\hat{H}\hat{Q}\left(E-\hat{H}\right)^{-1} 
\hat{Q} \hat{H}\hat{P}|\varphi \rangle
= E |\varphi \rangle\; .
\label{helpphi}
\end{equation}
We now set $\hat{P}\equiv \hat{P}_r$ so that
\begin{equation}
\hat{P}_r\hat{H}\hat{P}_r = \hat{T} + \hat{P}_r\hat{W}\hat{P}_r\;, \;
\hat{P}_r\hat{H}\hat{Q}_r = \hat{P}_r\hat{W}\hat{Q}_r\;,\;
\hat{Q}_r\hat{H}\hat{P}_r=\hat{Q}_r\hat{W}\hat{P}_r \, ,
\end{equation}
and insert the unit operator in~(\ref{helpphi}) to obtain
\begin{equation}
\hat{P}_r\hat{H}\hat{P}_r|\varphi\rangle+ 
\sum_{|m\rangle}\frac{\hat{P}_r\hat{W}\hat{Q}_r|m\rangle 
\langle m|\hat{Q}_r \hat{W}_r\hat{P}_r|\varphi\rangle}{E_{\varphi}-E_m}
= E_{\varphi}|\varphi\rangle\;.
\label{pt_downfolding}
\end{equation}
The same expression is obtained in Brillouin--Wigner perturbation theory
in which the energy $E\equiv E_{\varphi}$ must be determined self-consistently.

\subsubsection{Size-consistent reformulation.}

The down-folding or Brillouin--Wigner perturbation expansion 
is hampered by the fact that it is not size-consistent.
Therefore, the form~(\ref{pt_downfolding})
cannot be used for our calculations.

In order to make progress, we first subtract the ground-state energy
from all energies in~(\ref{pt_downfolding}),
\begin{equation}
(E_{\varphi}-E_0)|\varphi\rangle=
\hat{P}_r (\hat{H} -E_0) |\varphi\rangle +
\sum_{|m\rangle}\frac{\hat{P}_r\hat{W}\hat{Q}_r|m\rangle \langle m
|\hat{Q}_r \hat{W}|\varphi\rangle}{(E_{\varphi}-E_0)-(E_m-E_0)}\;,
\label{pt_df_luecke}
\end{equation}
so that all energy differences correspond to excitation energies 
of order unity. To second order in the interaction we may write
\begin{equation}
E_{\varphi}-E_0 \equiv e_{\varphi}\quad, \quad
E_m-E_0 = E_m^{(0)}-E_0^{(0)} + {\cal O}(U,V)\; .
\end{equation}
Therefore, to second order we find
\begin{equation}
e_{\varphi} |\varphi\rangle = 
\hat{P}_r (\hat{H} -E_0)|\varphi \rangle
+\sum_{|m\rangle}\frac{\hat{P}_r\hat{W}\hat{Q}_r|m\rangle 
\langle m|\hat{Q}_r \hat{W}|\varphi\rangle}%
{e_{\varphi}-(E_m^{(0)}-E_0^{(0)})}\; .
\label{bwdfsizeconsistent}
\end{equation}
As an example, we apply this equation for the ground state, i.e.,
$r=0$, $|\varphi\rangle=|{\rm FS}\rangle$, and $e_{\varphi}=0$.
When we multiply~(\ref{bwdfsizeconsistent})
from the left with $\langle {\rm FS} |$ we immediately recover 
the result from Rayleigh--Schr\"odinger perturbation theory.

For the exciton states~$|x\rangle$ 
and energies~$e_x$ we find the Schr\"odinger equation
\begin{equation}
e_x |x \rangle = \hat{P}_1 (\hat{H} -E_0)|x \rangle
+\sum_{|m\rangle}\frac{\hat{P}_1\hat{W}\hat{Q}_1|m\rangle 
\langle m|\hat{Q}_1\hat{W} |x\rangle}{e_x-(E_m^{(0)}-E_0^{(0)})}\; .
\end{equation}
This translates into a matrix-diagonalization problem with the diagonal
entries
\begin{eqnarray}
H_{\rm BW}(e_{(s,t)};k_1,k_1)&=&
{}_{(s,t)}\langle k_1|\hat{T}|k_1\rangle_{(s,t)}
-\langle {\rm FS}|\hat{T}|{\rm FS}\rangle
 -E_0^{(2)} 
\nonumber\\
&&
+{}_{(s,t)}\langle k_1|\hat{W}|k_1\rangle_{(s,t)}
-\langle {\rm FS}|\hat{W}|{\rm FS}\rangle \nonumber \\
&&
+\sum_{|m\rangle}\frac{{}_{(s,t)}\langle k_1|\hat{W}|m\rangle_{(s,t)} 
{}_{(s,t)}\langle m|\hat{W}|k_1\rangle_{(s,t)}}%
{e_{(s,t)}-(E_m^{(0)}-E_0^{(0)})}\; ,
\label{BWfirst}
\end{eqnarray}
and the non-diagonal entries ($k_1\neq k_2$)
\begin{eqnarray}
H_{\rm BW}(e_{(s,t)};k_1,k_2)&=&
{}_{(s,t)}\langle k_1|\hat{W}|k_2\rangle_{(s,t)}
\nonumber\\
&&
+\sum_{|m\rangle}\frac{{}_{(s,t)}\langle k_1|\hat{W} |m\rangle_{(s,t)} 
{}_{(s,t)}\langle m|\hat{W}|k_2\rangle_{(s,t)}}%
{e_{(s,t)}-(E_m^{(0)}-E_0^{(0)})}\; ,\nonumber \\
&& \label{heff_mat}
\end{eqnarray}
where the excited states~$|m\rangle$ contain none, two, or 
three particle-hole excitations of the Peierls ground state. Note that
the eigenvalue $e_{(s,t)}$ of the matrix $H_{\rm BW}(e_{(s,t)})$
in~(\ref{heff_mat}) must be determined self-consistently.
In our calculations we target the lowest eigenvalue as the bound exciton in
the triplet and singlet sectors.

\section{Excitons to second order}
\label{sec:excitons2nd}

\subsection{Analytical results}

The calculation of optical excitation energies 
requires the matrix elements
\begin{equation}
M_{0, (s,t)}(p_1,p_2) = {}_{(s,t)}\langle p_1|\hat{T}|p_2\rangle_{(s,t)}
- \delta_{p_1,p_2}E_0^{(0)}=  \delta_{p_1,p_2}2E(p_1) \; ,\label{m0}
\end{equation}
and 
\begin{eqnarray}
M_{1, (s,t)}(p_1,p_2) &=& {}_{(s,t)}\langle p_1|\hat{W}|p_2\rangle_{(s,t)}
- \delta_{p_1,p_2}E_0^{(1)}\; ,\label{m1}\\
M_{2, (s,t)}(e;p_1,p_2) &=& \sum_{|m\rangle}
\frac{{}_{(s,t)}\langle p_1|\hat{W} |m\rangle_{(s,t)} 
{}_{(s,t)}\langle m|\hat{W}|p_2\rangle_{(s,t)}}%
{e-(E_m^{(0)}-E_0^{(0)})}\; ,\label{m2}
\end{eqnarray}
where $|m\rangle$ contains up to three particle-hole excitations.

\subsubsection{Matrix elements to first order.}

The matrix elements $M_1$ are readily calculated.
We find
\begin{equation}
M_{1, (s,t)}(p_1,p_2)  = M_{1, (s,t)}^U(p_1,p_2) + M_{1, (s,t)}^V(p_1,p_2) 
\label{opt_2}
\end{equation}
with
\begin{equation}
M_{1, (s,t)}^U(p_1,p_2)  = \pm \frac{U}{L}\; ,
\label{opt_u}
\end{equation}
and 
\begin{eqnarray}
M_{1, (s,t)}^V(p_1,p_2)  &=& \delta_{p_1,p_2}
\sum_{-L/2<r<L/2 \atop{r|2=1}} V(r)
\Bigl[\frac{8\cos(p_1)\cos(p_1r)}{E(p_1)}A_{\delta}(r) \nonumber \\
&& \hphantom{\frac{1}{2}\sum_{-L/2<r<L/2 \atop{r|2=1}} V(r)\delta_{p_1,p_2} }
+ \frac{8\delta\sin(p_1)\sin(p_1r)}{E(p_1)}B_{\delta}(r)\Bigr]
\nonumber\\
&&-\frac{2}{L}\sum_{r\neq 0}
V(r)\cos((p_1-p_2)r)\nonumber \\
&&\hphantom{-\frac{2}{L}\sum_{r\neq 0}}
\left[f_2^2(p_1,p_2)+(-1)^rf_1^2(p_1,p_2)\right]
\nonumber\\
&&+\frac{2}{L}\sum_{r\neq 0}V(r)\left[1+(1\pm1)(-1)^r\right]\; .
\label{opt_v}
\end{eqnarray}

\subsubsection{Matrix elements to second order.}

The intermediate states
$|00\rangle=|{\rm FS}\rangle$,
$|11\rangle$, $|20\rangle$, $|02\rangle$,
$|30\rangle$, $|03\rangle$, $|21\rangle$, and $|12\rangle$
contribute to~(\ref{m2}). The respective terms are lengthy, see~\ref{appB}.
As in sections~\ref{sec:gsener} and~\ref{sec:singleP} we write
\begin{eqnarray}
M_{2, (s,t)}(e;p_1,p_2) &=& M_{2,(s,t)}^{U^2}(e;p_1,p_2) 
+  M_{2,(s,t)}^{V^2}(e;p_1,p_2) \nonumber \\
&& 
+ M_{2, (s,t)}^{UV}(e;p_1,p_2)\; . 
\end{eqnarray}
The contributions to $M_{2,(s,t)}^{U^2}(e;p_1,p_2)$ can be found 
in~(\ref{opt_u2_ztla11}) and~(\ref{opt_u2_dtla21}).
The terms for $M_{2,(s,t)}^{V^2}(e;p_1,p_2)$ result 
from~(\ref{Vsquare00}), (\ref{Vsquare11}), (\ref{Vsquare20}),
(\ref{Vsquare21}), and~(\ref{Vsquare30}).
Lastly, the terms for $M_{2,(s,t)}^{UV}(e;p_1,p_2)$ 
are collected in~(\ref{UV11}) and~(\ref{UV21}).

\subsubsection{Wannier perturbation theory.}

The first-order result from Wannier perturbation theory for
the singlet/triplet excitons follows from the lowest eigenvalue 
$e_{(s,t)}^{(1)}$ of
the Wannier matrix with entries
\begin{equation}
M^{\rm W}(p_1,p_2) = M_{0, (s,t)}(p_1,p_2)+M_{1, (s,t)}(p_1,p_2)\; .
\end{equation}
Moreover, this calculation gives the first-order Wannier wave functions
with (real) amplitudes $w_{(s,t)}(p)$, cf.~(\ref{exz_wf}).

To second order, we evaluate~(\ref{2O}),
\begin{equation}
e_{(s,t)}^{(2)}=  \sum_{|m\rangle}
\frac{\langle (s,t)|\hat{H}_{\perp}|m\rangle
\langle m|\hat{H}_{\perp}|(s,t)\rangle}%
{e_{(s,t)}^{(0)}-(E_m^{(0)}-E_0^{(0)})}\; .
\label{wpt_2o}
\end{equation}
On the right-hand-side of this equation
we keep only the kinetic energy of the Wannier exciton,
\begin{equation}
e_{(s,t)}^{(0)} = \sum_{k} 2E(k) [w_{(s,t)}(k)]^2
\label{ezeroWA}
\end{equation}
in order to be consistent to second order in the Coulomb interaction.
The second-order correction of the Wannier excitons' excitation energy
is then given by
\begin{equation}
e_{(s,t)}^{(2)} = \sum_{p_1,p_2} 
w_{(s,t)}(p_1)w_{(s,t)}(p_2) M_{2, (s,t)}(e_{(s,t)}^{(0)};p_1,p_2)-E_0^{(2)}
\; ,
\end{equation}
and the excitons' excitation 
energy is $e_{(s,t)}=e_{(s,t)}^{(1)} + e_{(s,t)}^{(2)}$.

\subsubsection{Two-step perturbation theory.}

To first order, the results from the Wannier and 
two-step perturbation theories are identical. For the calculation
of the corrections we define the matrix
\begin{eqnarray}
B_{(s,t)}(p_1,p_2) &=&  \sum_{|m\rangle}
{}_{(s,t)}\langle p_1|\hat{W} |m\rangle \langle m|\hat{W}|p_2\rangle_{(s,t)}
\nonumber\\
&& \frac{E_m^{(0)}-E_0^{(0)}}%
{(2E(p_1)-(E_m^{(0)}-E_0^{(0)}))(2E(p_2)-(E_m^{(0)}-E_0^{(0)}))}\; .\nonumber\\
&& 
\end{eqnarray}
In order to calculate the entries of this matrix the energy denominators
which appear in~\ref{appB} must be modified appropriately.

As shown in~(\ref{twostep}),
the excitation energy of the excitons in two-step perturbation theory 
is then obtained from the lowest eigenvalues of the matrix with the entries
\begin{eqnarray}
A_{(s,t)}(p_1,p_2) &=& M_{0, (s,t)}(p_1,p_2)+M_{1, (s,t)}(p_1,p_2) \nonumber \\
&& +\delta_{p_1,p_2}\left( M_{2, (s,t)}(2E(p_1);p_1,p_1)-E_0^{(2)}\right)
\nonumber\\
&&+ (1-\delta_{p_1,p_2})\Bigl( M_{2, (s,t)}(2E(p_1);p_1,p_2) \nonumber \\
&&\hphantom{+(1-\delta_{p_1,p_2})\Bigl( }
+M_{2, (s,t)}(2E(p_2);p_1,p_2) \nonumber\\
&&\hphantom{+(1-\delta_{p_1,p_2})\Bigl( }
+B_{(s,t)}(p_1,p_2)\Bigr)\; .
\end{eqnarray}

\subsubsection{Down-folding perturbation theory.}

According to~(\ref{BWfirst}) and (\ref{heff_mat}) the down-folding
method requires the diagonalization of the matrix with the
entries
\begin{eqnarray}
F_{(s,t)}(e_{(s,t)};p_1,p_2) &=& M_{0, (s,t)}(p_1,p_2)+M_{1,(s,t)}(p_1,p_2)
\nonumber  \\
&& +M_{2, (s,t)}(e_{(s,t)};p_1,p_2)-\delta_{p_1,p_2}E_0^{(2)}\; .
\end{eqnarray}
The lowest eigenvalue must be determined self-consistently.

\subsubsection{Comparison of numerical effort.}

The numerically cheapest method is the two-step perturbation theory.
For given Peierls dimerization~$\delta$ and functional form of~$V(r)$
one can calculate and store the matrix elements which 
define~$A_{(s,t)}(p_1,p_2)$. The parameters $U$, $V$ only enter
as multiplicative factors.

This does not hold for the Wannier approach anymore because
the kinetic energy depends on the wave function of the Wannier
exciton in first order of the interaction.
For a first guess we could replace $e_{(s,t)}^{(0)}$
in~(\ref{ezeroWA}) by $2E(-\pi/2)$ because $p=-\pi/2$ is the
dominant contribution in the excitonic wave functions.
This speeds up the analysis as a function of the Coulomb parameters
but it also reduces the quality of the results.

The down-folding approach is the most costly of the three approaches
in terms of computer-time.
For given interaction parameters, the effort of the
second-order Wannier theory has to be repeated some six to eight times
until convergence is reached.

\subsection{Comparison with numerical results for the singlet exciton}

We start with a comparison of our analytical results 
for the singlet exciton with those of the DMRG. For the latter we assume
that the singlet exciton is identical to the lowest-lying charge
excitation at fixed particle number $N=L$. 
This assumption is justified in the presence of a singlet exciton.

We do not present results from the down-folding approach.
The self-consistency algorithm is stable but the results from
the down-folding approach underestimate the excitonic excitation
energies drastically. We do not consider the down-folding
perturbation theory any further because it is numerically the most
expensive and quantitatively the least successful of our perturbative
approaches.

\subsubsection{Finite-size effects.}

\begin{figure}[htbp]
\begin{center}
\includegraphics[width=11.8cm]{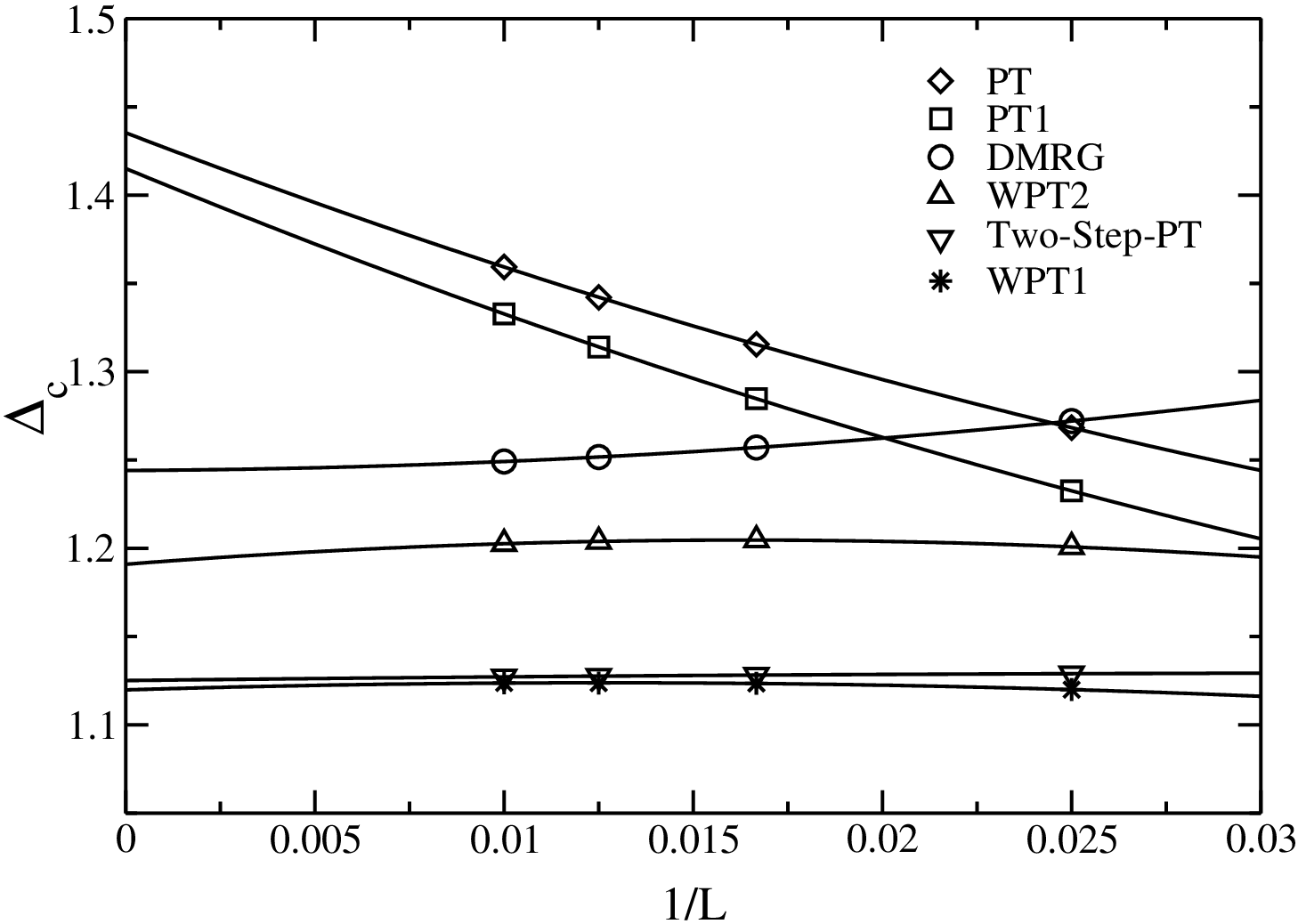}
\caption{Singlet exciton energy (charge gap)
$\Delta_c$ as a function of inverse system size for fixed $U=2V=2t$
  and $\delta=0.2$ in the DMRG and perturbation theory; 
see text for further details.\label{Fig:cgap-vs-l}}
\end{center}
\end{figure}

As in previous sections we first investigate the size-dependence
of the excitation energy. In Fig.~\ref{Fig:cgap-vs-l}
we compare the results from the DMRG with those from the various
perturbative approaches which we discussed in 
section~\ref{sec:ptapproachestoexc}. 
The DMRG data show a very weak size dependence which is
a clear signal for a bound state. The data PT and PT1
do not reproduce this $1/L$ dependence. In fact, they describe
an {\sl unbound\/} particle-hole excitation. 
The starting point for PT and PT1 is a single particle-hole excitation 
at momentum $p=-\pi/2$, and corrections are calculated
in first and second-order Rayleigh--Schr\"odinger perturbation theory,
respectively. The results show that this approach is not applicable
for the description of excitonic optical excitations, and we shall
not pursue this approach any further.

Wannier theory to first order (WPT1), to second order (WPT2), 
and the two-step perturbation theory show a very weak size dependence,
characteristic for a bound state. As seen from the figure,
the two-step perturbation theory does not improve the result
from first-order Wannier perturbation theory, and the 
second-order Wannier theory is the best approximation to the
DMRG data. 

\begin{figure}[htbp]
\begin{center}
\includegraphics[width=12cm]{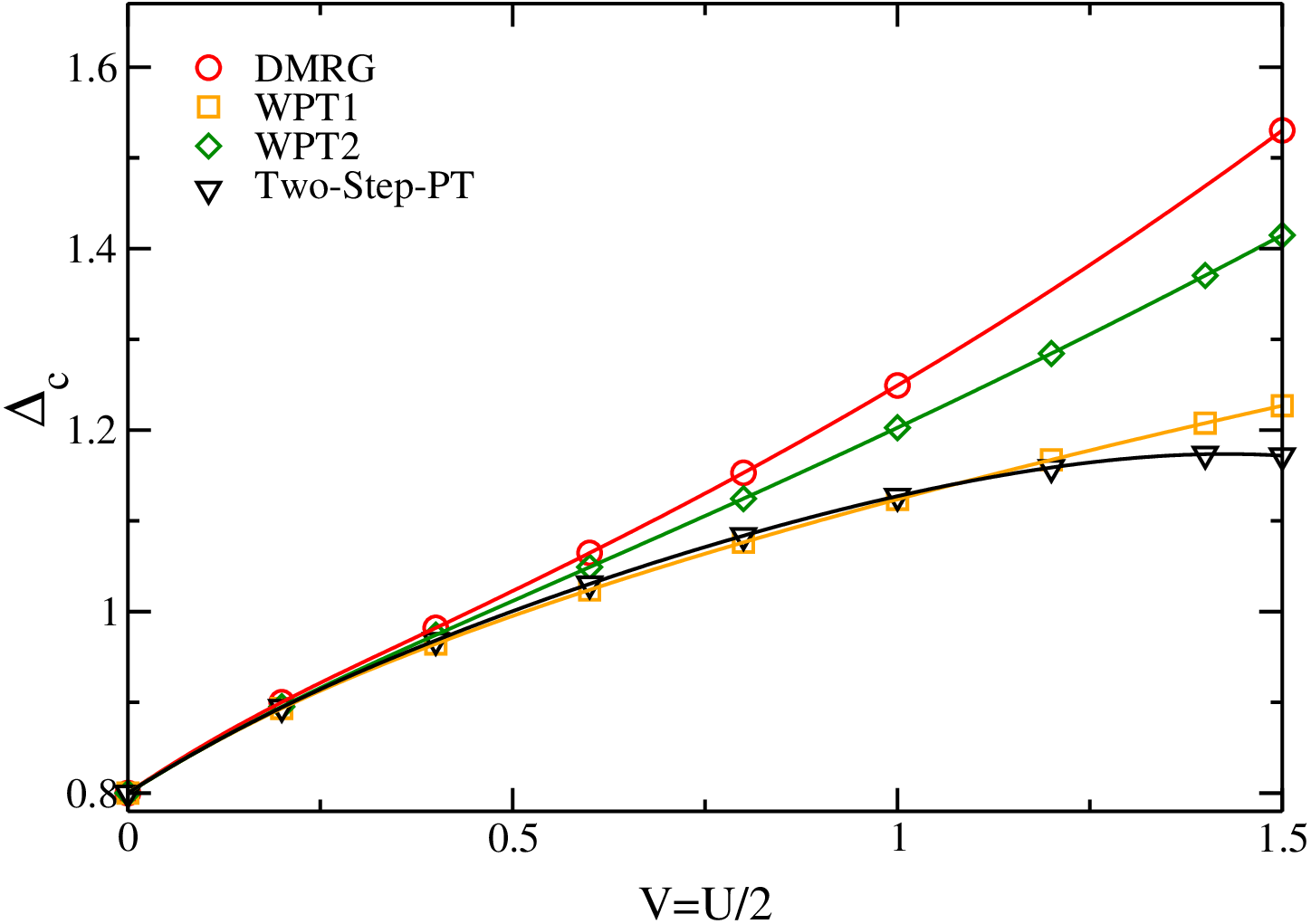}
\caption{Singlet exciton energy (charge gap)
$\Delta_c$ as a function of~$V$
for fixed ratio $U/V=2$, $L=100$, and $\delta=0.2$ in the DMRG and in various
perturbative approaches.\label{Fig:cgap-vs-dmrg-u2v}}
\end{center}
\end{figure}

In all cases, the results for $L=100$ are a very good estimate
for the results in the thermodynamic limit. 
The differences between the various perturbative approaches
and the numerically exact DMRG are significant.

\subsubsection{Fixed ratio $U/V$.}

In Fig.~\ref{Fig:cgap-vs-dmrg-u2v} we show the results 
for the singlet excitation energy as a function of~$V$
for $L=100$, $\delta=0.2$, and fixed ratio $U/V=2$.
It is seen that second-order Wannier theory is naturally superior
to the first order approximation and provides a rather good
description. At ($V=1.5t$, $U=3t$), 
when the charge gap has almost doubled from its
non-interacting value $\Delta^{\rm P}$, the difference
between the DMRG and second-order Wannier theory is less than ten percent.
In contrast, the first-order Wannier theory is already off by
30~percent, and two-step perturbation theory becomes even worse than
that. 

\subsubsection{Fixed~$V$.}

\begin{figure}[htbp]
\begin{center}
\includegraphics[width=11cm]{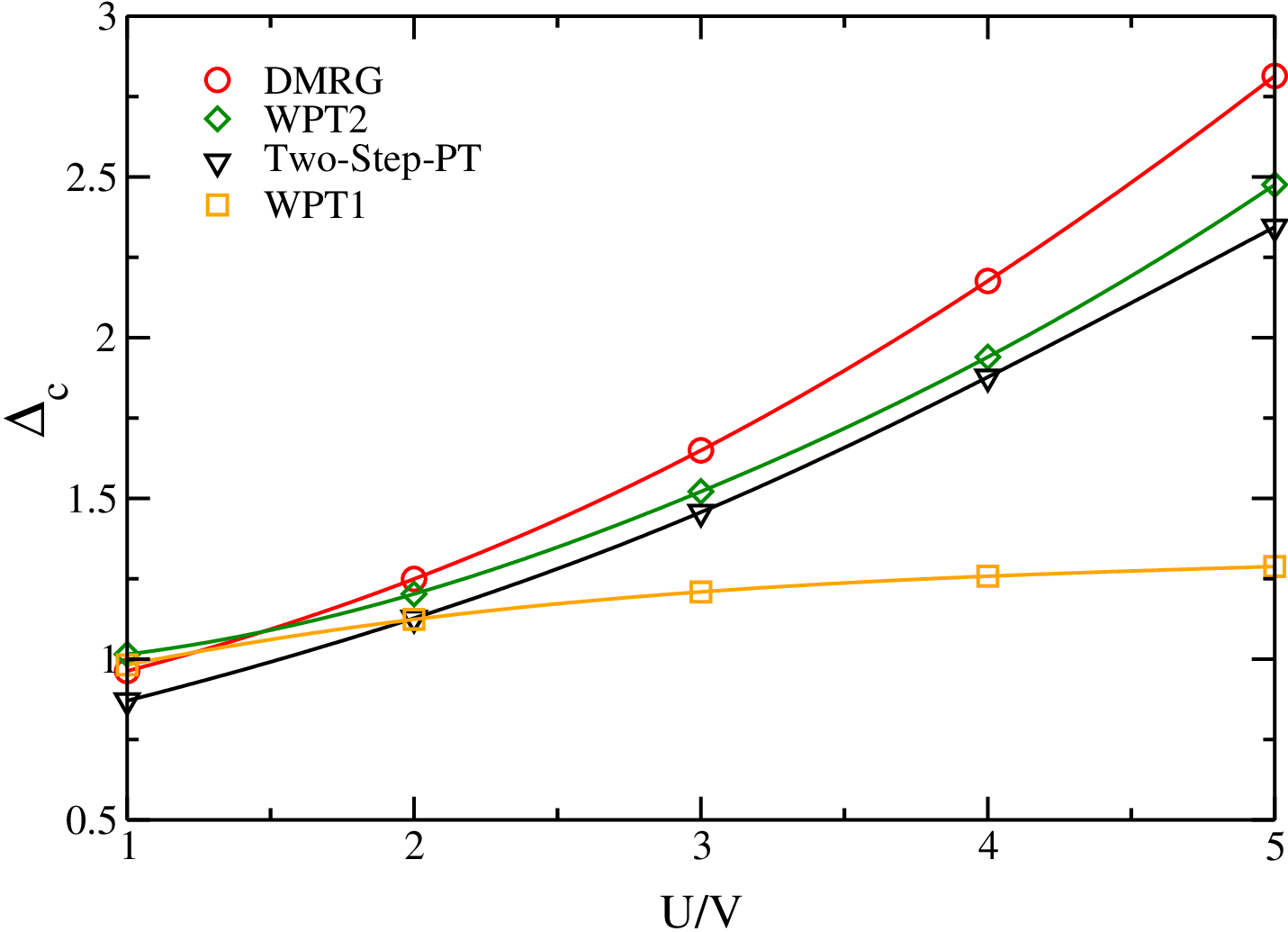}
\caption{Singlet exciton energy (charge gap)
$\Delta_c$ as a function of~$U/V$ for fixed $V=t$, $L=100$, and 
$\delta=0.2$ in the DMRG and in various perturbative 
approaches.\label{Fig:cgap-vs-dmrg-v1}}
\end{center}
\end{figure}

As for the ground-state energy and the single-particle gap,
the discrepancies between the DMRG 
and second-order perturbation theory increase
as a function of~$V$ but not so much as a function of~$U$.
In Fig.~\ref{Fig:cgap-vs-dmrg-v1} we show the singlet excitation
energy as a function of~$U$ for fixed $V=t$, $L=100$, and
$\delta=0.2$. The two-step perturbation theory and second-order
Wannier theory closely follow the almost quadratical increase
of the charge gap as a function of~$U$. The singlet excitation
energy can triple from its bare value~$\Delta^{\rm P}=0.8t$
at ($U=5t$, $V=t$), and still the deviations are only about
ten percent. In contrast,
first-order Wannier theory begins to deviate noticeably
already at ($U=3t, V=t$). 

\subsubsection{Conclusion.}

Wannier theory to second order consistently improves the results from
first-order Wannier theory. It provides a quantitative description
even for large Coulomb parameters when the charge gap has more than
doubled from its bare Peierls value. 

The two-step perturbation theory can handle the influence of large
Hubbard interactions but it fails quickly when the long-range
parts of the Coulomb interactions become relevant. 
This is somewhat unfortunate because the two-step perturbation
theory is numerically the cheapest of all second-order methods.

The good performance of the second-order Wannier theory
indicates that this approach describes the relevant objects
appropriately. To first order in the
Coulomb interaction Wannier theory forms the charge neutral
exciton from a negative (electron) and a positive charge excitation
(hole). This object has the appropriate quantum numbers also
for strong Coulomb interactions. 
Therefore, the residual Coulomb interactions,
i.e, the coupling to the ground state and other particle-hole
excitations, apparently perturb the exciton properly.
In contrast, two-step perturbation theory builds excitons 
from dressed electrons and holes and thereby overestimates
the repulsion against other states with electron-hole pairs
which results in a too small energetic splitting between the 
exciton in two-step perturbation theory and the ground state.

\subsection{Comparison with numerical results for the triplet exciton}

We continue with a comparison of our analytical results 
for the triplet exciton with those of the DMRG. For the latter we assume
that the triplet exciton is identical to the lowest-lying spin
excitation at fixed particle number $N=L$. 
This assumption is justified in the presence of a triplet exciton.
Again, we do not present results from the down-folding approach
because it fails quantitatively already for fairly small interaction
strengths.

\subsubsection{Fixed ratio $U/V$.}

\begin{figure}[htbp]
\begin{center}
\includegraphics[width=12cm]{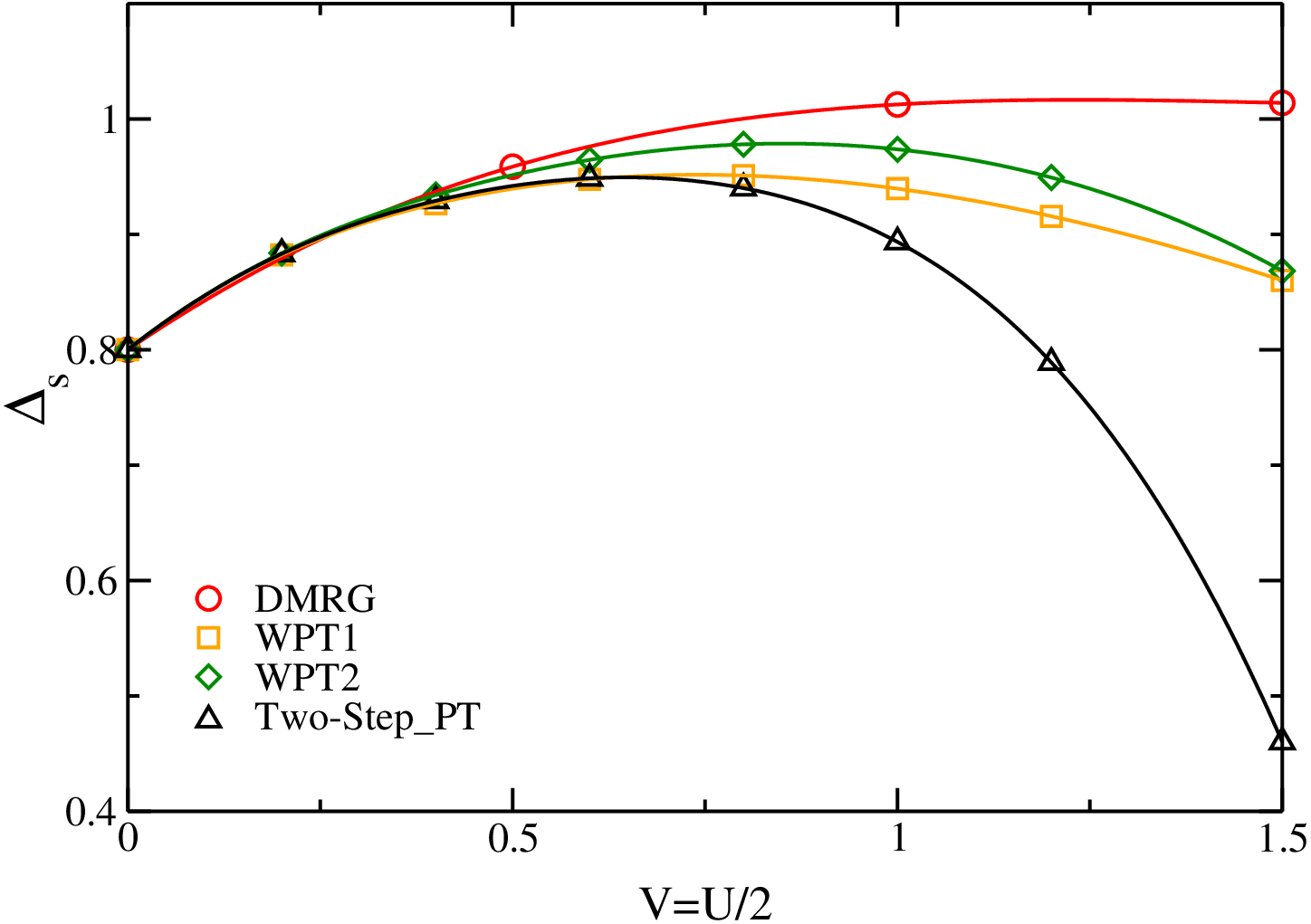}
\caption{Triplet exciton energy (spin gap) $\Delta_s$ as a function of~$V$
for fixed ratio $U/V=2$, $L=100$, and $\delta=0.2$ 
in the DMRG and in various
perturbative approaches.\label{Fig:sgap-vs-dmrg-u2v}}
\end{center}
\end{figure}

For the triplet exciton, the results from perturbation theory are much
less reliable than for the singlet exciton. As can be seen from 
Fig.~\ref{Fig:sgap-vs-dmrg-u2v}, a quantitatively correct answer is
provided by second-order Wannier theory only up to $U=2V=2t$,
whereas first-order Wannier theory and two-step perturbation theory
fail for even smaller Coulomb parameters. 

\subsubsection{Fixed~$V$.}

\begin{figure}[htbp]
\begin{center}
\includegraphics[width=11cm]{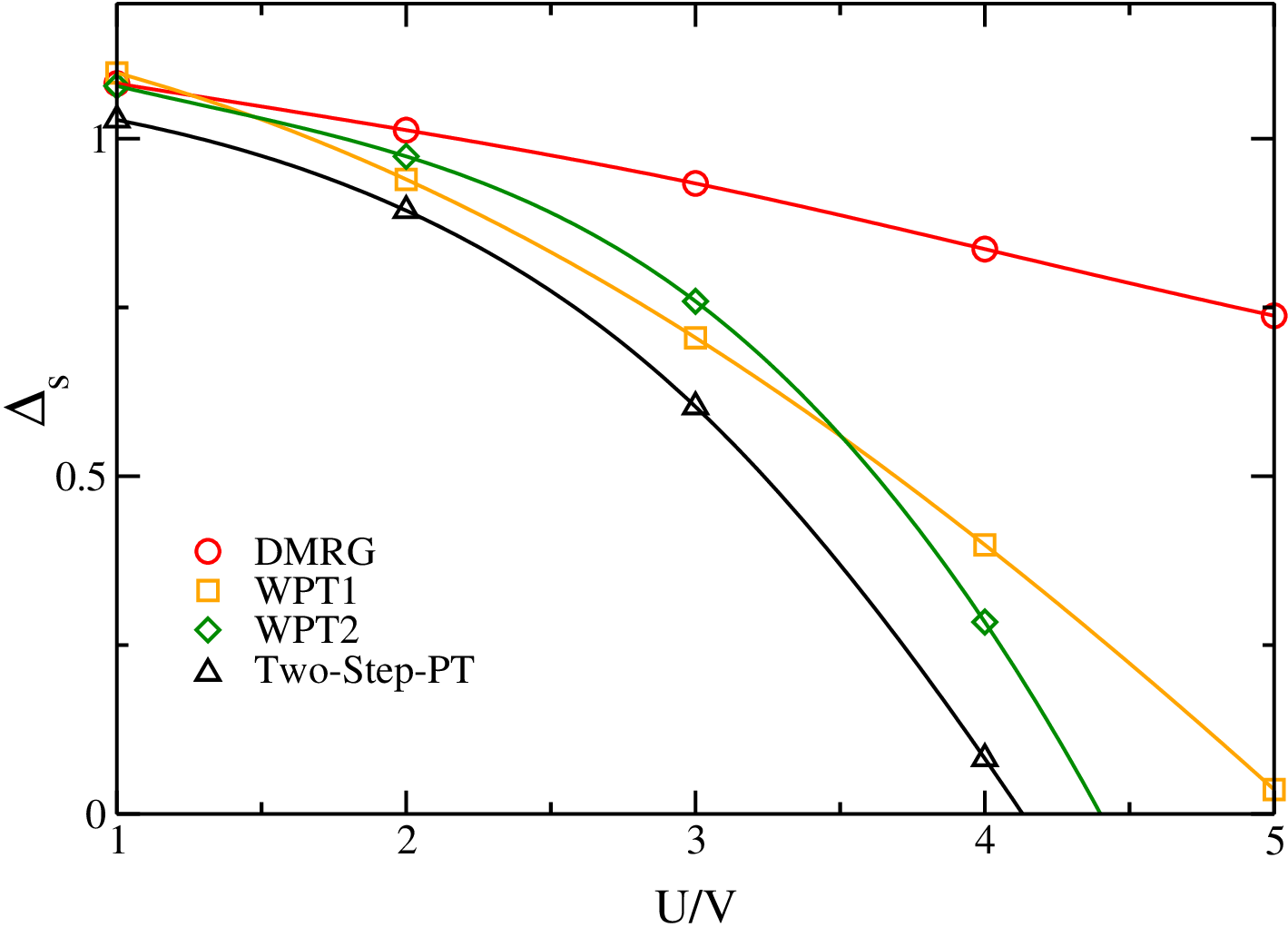}
\caption{Triplet exciton energy (spin gap) $\Delta_s$ as a function
  of~$U/V$
for fixed $V=t$, $L=100$, and $\delta=0.2$ in the DMRG 
and in various perturbative approaches.\label{Fig:sgap-vs-dmrg-v1}}
\end{center}
\end{figure}

Perturbation theory fails even qualitatively for moderate to large
values of the Hubbard interaction. 
The DMRG exciton energy increases, flattens out, 
and finally turns into a $1/U$ behavior, as seen in 
Figs.~\ref{Fig:sgap-vs-dmrg-u2v} and~\ref{Fig:sgap-vs-dmrg-v1}.
This shows that the nature of the excitonic excitation changes
drastically as a function of~$U/t$: for large Coulomb interactions
the triplet exciton is a bound state of a pair of two spins
on singly occupied sites, i.e., it looses its character
of bound charge excitations which is assumed in perturbation theory. 

\subsubsection{Conclusion.}

Triplet excitons cannot be described reliably within 
low-order perturbation theory because the nature of
the lowest triplet excitation quickly changes as 
a function of the interaction strength. For not too large
Hubbard interactions, the triplet exciton looses its
charge contents and is better viewed as a bound state
of two spin-1/2 excitations (spinons).
The description of such a state lies beyond 
the possibilities of finite-order perturbation theory.

\section{Concluding remarks}
\label{sec:discussion}

\subsection{Summary}

In this work we performed second-order perturbation theory for 
the extended Peierls--Hubbard model for the ground-state energy
and the single-particle gap. The agreement with the numerically exact 
DMRG data was very good up to Coulomb parameters 
where the bare Peierls gap has almost tripled.

For bound optical excitations, the situation is less favorable.
The singlet exciton energy is reasonably well described
by second-order Wannier perturbation theory for Coulomb parameters
which lead to a doubling of the single-particle gap. The triplet
exciton can be described perturbatively for weak coupling only.

The second-order Wannier theory proves to be superior to
other second-order approaches. Its success may be related to the fact
that after its formation the exciton is a charge neutral object
to which other parts of the Hilbert space are coupled less intensely
by the Coulomb interaction.
The computationally most
costly down-folding method performs worst and can be safely discarded.
The two-step perturbation theory where excitons are formed from
dressed particles is computationally the cheapest method.
Unfortunately,
it does not improve the first-order Wannier theory for the singlet exciton
systematically. This indicates that 
the exciton should not be seen as a bound state of dressed
quasi-particles. In order to investigate this question further,
the GW-BSE approach~\cite{GWBSE} ought to be applied directly 
to the extended Peierls--Hubbard model so that its predictions
can be tested against the numerically exact DMRG results.

\subsection{Discussion}

In order to compare our results with those of previous work on 
poly-(di)acetylene, we note that we may write
$V=e^2/(\varepsilon a_0)$ where $a_0\approx 1.4$\AA~is 
the average distance between the carbon atoms, $e$~is the electric
charge and $\varepsilon$ is the dielectric constant of the surrounding
medium. In~\cite{Abe} its value is given by $\varepsilon_{\rm Abe}\approx 5$
which is much larger than the typical values for plastic,
$\varepsilon\approx 2.3$. For the Abe value one finds
$V\approx 2\, {\rm eV}=t$, and the application of first-order
(Wannier) perturbation theory at $\delta=0.2 $ gives a reasonable
agreement with the measured
binding energies of the first singlet and triplet exciton
in poly-diacetylene which are $\Delta_{\rm c}=0.5\, {\rm eV}$ and 
$\Delta_{\rm s}=0.9\, {\rm eV}$, respectively.
However, the measured single-particle gap is
$\Delta_1=2.3\, {\rm eV}$ which is seriously overestimated 
by $0.7\, {\rm eV}$ for the given choice of parameters, 
$\Delta_1^{\rm DMRG}(U=2t,V=t,\delta=0.2)=3.0\,{\rm eV}$.

The reason for this discrepancy is readily identified.
The assumption $\delta=0.2$ leads to a bare gap of 
$\Delta_{\rm P}=0.8 t\approx 1.6\,{\rm eV}$. 
As has been seen in Fig.~\ref{Fig:etl-energy-vs-dmrg}, the
single-particle gap increases linearly with the Coulomb interaction.
Therefore, a good agreement with the experimental gap value is achieved
at $V=0.6 t$ which would correspond to $\varepsilon\approx 8$
which is unrealistic. Apparently, the assumption of a large
Peierls gap is not correct. 

A more realistic approach starts from a much smaller Peierls
contribution to the gap. We should rather work with $\delta=0.1$
or smaller~\cite{Lavrentiev}, 
which is in accordance with an estimate of the electron-lattice
interaction in benzene~\cite{Kiess}.
The Peierls gap then is $\Delta_{\rm P}
=0.4 t \approx 0.8\, {\rm eV}$ which is about one third of the full
gap. We then find for the single-particle gap
$\Delta_1^{\rm PT}(U=2t,V=t,\delta=0.1)=1.75\,{\rm eV}$
from second-order perturbation theory which is 12~percent {\sl below\/} the
exact value of $\Delta_1^{\rm DMRG}(U=2t,V=t,\delta=0.1)=1.98\,{\rm eV}$
from the DMRG.
For the singlet exciton, we find 
$\Delta_{\rm c}^{\rm PT}(U=2t,V=t,\delta=0.1)=1.44\,{\rm eV}$
which is 22~percent {\sl above\/} the DMRG value
$\Delta_{\rm c}^{\rm DMRG}(U=2t,V=t,\delta=0.1)=1.18\,{\rm eV}$.
The agreement is quite acceptable in both cases.
However, the comparison of the binding energies for the singlet exciton,
$\Delta_{\rm bind, s}^{\rm PT}(U=2t,V=t,\delta=0.1)=0.31\,{\rm eV}$
versus $\Delta_{\rm bind, s}^{\rm DMRG}(U=2t,V=t,\delta=0.1)=0.80\,{\rm eV}$
reveals that we are stretching second-order perturbation theory 
beyond its limits
because the error in the binding energy increases to more than 60~percent.

A realistic description of the experimental data for poly-diacetylene 
requires the inclusion of the full band-structure 
and of the lattice relaxation for excited states, mostly for
the triplet exciton~\cite{Barford}. In particular,
a better understanding of the screening in these materials~\cite{BBYaron}
and a refinement of the Ohno potential are necessary. 

\ack

We thank H.~Benthien and R.M.~Noack for helpful discussions.
J.~Rissler acknowledges support by the Alexander-von-Humboldt Foundation.
\appendix

\section{Useful relations}
\label{appA}

\subsection{Peierls transformation}

The Peierls Hamiltonian is diagonalized via the transformation 
\begin{equation}
\hat{a}_{k,\sigma}  = \alpha_k\hat{c}_{k,\sigma}
+\mathi \beta_k\hat{c}_{k+\pi,\sigma}\; , \quad
\hat{b}_{k,\sigma} = \beta_k\hat{c}_{k,\sigma}
-\mathi \alpha_k\hat{c}_{k+\pi,\sigma}\; .
\end{equation}
Its inverse reads
\begin{equation}
\hat{c}_{k,\sigma}  =  \alpha_k\hat{a}_{k,\sigma}
+\beta_k\hat{b}_{k,\sigma}\;, quad
\hat{c}_{k+\pi,\sigma}  =  -\mathi \beta_k\hat{a}_{k,\sigma}
+\mathi \alpha_k\hat{a}_{k,\sigma}\; .
\end{equation}
Moreover, we have
\begin{equation}
2\alpha_k\beta_k = -\frac{2t\delta \sin(k)}{E(k)}\;, \quad 
\alpha_k^2-\beta_k^2 = \frac{2t\cos(k)}{E(k)}\; .
\label{2ab_cos}
\end{equation}

\subsection{Help functions}

In order to express matrix second-order elements for the
Hubbard interaction we
introduce the abbreviations
\begin{eqnarray}
f_1(k_1,k_2) & \equiv & \alpha_{k_1}\beta_{k_2}- \alpha_{k_2}\beta_{k_1} 
= -f_1(k_2,k_1) \; , \nonumber \\
f_2(k_1,k_2) & \equiv & \alpha_{k_1}\alpha_{k_2}+ \beta_{k_1}\beta_{k_2} 
= f_2(k_2,k_1) \; ,\nonumber \\
u_1(k_1,k_2,k_3,k_4) &\equiv& f_1(k_2,k_1)f_1(k_4,k_3)-f_2(k_2,k_1)f_2(k_4,k_3)
\;, \nonumber \\
u_2(k_1,k_2,k_3,k_4) &\equiv& f_1(k_2,k_1)f_2(k_4,k_3)+f_1(k_4,k_3)f_2(k_2,k_1)
\; .
\label{uabbr}
\end{eqnarray}
The second-order matrix elements for the long-range Coulomb 
require the functions
\begin{eqnarray}
A_\delta(r) & \equiv &\frac{1}{L}\sum_k
\frac{\cos(k)\cos(kr)}{\sqrt{\cos(k)^2+\delta^2\sin(k)^2}}
=A_\delta(-r)\;, \nonumber \\
B_\delta(r) & \equiv & \frac{1}{L}\sum_k
\frac{\delta\sin(k)\sin(kr)}{\sqrt{\cos(k)^2+\delta^2\sin(k)^2}}
=-B_\delta(-r)\; ,
\label{abdelta}\\
v_0(k) & \equiv & \sum_{-L/2<r<L/2,r|2=1} \frac{1}{2|r|}
\frac{4t}{E(k)}\Bigl(\delta\sin(k)\cos(kr)A_\delta(r) \nonumber \\[-6pt]
&& \hphantom{\sum_{-L/2<r<L/2,r|2=1} \frac{1}{2|r|}\frac{4t}{E(k)}\Bigl(}
-\cos(k)\sin(kr)B_\delta(r)\Bigr)\; , \label{v0}
\end{eqnarray}
where ($r|2=1$) denotes all odd~$r$.
With the help functions 
\begin{eqnarray}
C_+(k_1,k_2) & \equiv & \sum_{r\neq 0}\frac{1}{2|r|}2\cos(r(k_1-k_2))
\;, \nonumber \\
C_-(k_1,k_2) & \equiv & \sum_{r\neq 0}\frac{(-1)^r}{2|r|}2\cos(r(k_1-k_2)) 
\end{eqnarray} 
we then introduce 
\begin{eqnarray}
v_1(k_1,k_2,k_3,k_4) & \equiv & C_+(k_3,k_4)f_1(k_2,k_1)f_1(k_4,k_3) 
\nonumber \\
&& -C_-(k_3,k_4)f_2(k_2,k_1)f_2(k_4,k_3)\,; \label{v1} \\
v_2(k_1,k_2,k_3,k_4) & \equiv &C_+(k_3,k_4)f_1(k_4,k_3)f_2(k_2,k_1) \nonumber 
\\
&& +C_-(k_3,k_4)f_1(k_2,k_1)f_2(k_4,k_3) \label{v2} \; .
\end{eqnarray}

\subsection{Useful contractions}

We collect some contractions which frequently appear 
in the calculation of matrix elements to first and second order
in the Coulomb interaction.
The expression $\langle\hat{O}\rangle_0$
denotes the expectation value of the operator $\hat{O}$ in the
ground state~$|{\rm FS}\rangle$~(\ref{defFS}) of the Peierls insulator.
The pair contractions are
\begin{eqnarray}
\langle \hat{n}_{l,\sigma}\rangle_0 &=& \frac{1}{2}\;, \label{onehalf} \\
\langle \hat{c}^+_{l,\sigma}\hat{c}_{l+r,\sigma}\rangle_0 &=&
\frac{1}{2}\delta_{r,0}+\frac{1-(-1)^r}{2}(A_\delta(r)+(-1)^lB_\delta(r)) 
\;, \nonumber \\
\langle \hat{c}^+_{l,\sigma}\hat{a}_{k,\sigma}\rangle_0 &=&
\frac{1}{\sqrt{L}}e^{-\mathi kl}(\alpha_k+\mathi (-1)^l\beta_k)\;,\nonumber 
\\
\langle \hat{c}_{l,\sigma}\hat{b}^+_{k,\sigma}\rangle_0 &=&
\frac{1}{\sqrt{L}}e^{\mathi kl}(\beta_k+\mathi (-1)^l\alpha_k)\label{cb}\; .
\end{eqnarray}
Important contractions with four Fermi operators are
\begin{eqnarray}
\langle \hat{n}_{l,\sigma}\hat{b}^+_{k_1,\sigma}\hat{a}_{k_2,\sigma}\rangle_0 
&=&
\frac{1}{L}e^{\mathi (k_1-k_2)l}(f_1(k_2,k_1)
+\mathi (-1)^lf_2(k_2,k_1))\; ,\nonumber \\
\langle\hat{b}_{k,\sigma}\hat{c}^+_{l+r,\sigma}\rangle_0\langle
\hat{c}_{l,\sigma}\hat{b}^+_{k,\sigma}\rangle_0 &=&
\frac{1}{L}e^{-\mathi kr}(\beta_k^2+(-1)^r\alpha_k^2) \nonumber \\
&& 
+\mathi \frac{1}{L}e^{-\mathi kr}
(-1)^l(1-(-1)^r)\alpha_k\beta_k\;,\nonumber \\
\langle \hat{a}^+_{k,\sigma}\hat{c}_{l+r,\sigma}\rangle_0\langle
\hat{c}^+_{l,\sigma}\hat{a}_{k,\sigma}\rangle_0 &=&
-\frac{1}{L}e^{\mathi kr}(\alpha_k^2+(-1)^r\beta_k^2) \nonumber \\
&&
-\mathi \frac{1}{L}e^{\mathi kr}(-1)^l(1-(-1)^r)\alpha_k\beta_k\nonumber \; .
\end{eqnarray}
and
\begin{eqnarray}
\left\langle \hat{b}_{k_1,\sigma}
\left(\hat{n}_{l,\sigma}-\frac{1}{2}\right)
\hat{b}^+_{k_2,\sigma}\right\rangle_0 &=&
\frac{1}{L}e^{-\mathi (k_1-k_2)l}f_2(k_2,k_1) \nonumber \\
&& 
+\mathi \frac{1}{L}e^{-\mathi (k_1-k_2)l} (-1)^lf_1(k_2,k_1)\; ,  \\
\left\langle \hat{a}^+_{k_1,\sigma}
\left(\hat{n}_{l,\sigma}-\frac{1}{2}\right)
\hat{a}_{k_2,\sigma}\right\rangle_0 &=&
-\frac{1}{L}e^{\mathi (k_1-k_2)l}f_2(k_2,k_1) \nonumber \\
&& +\mathi \frac{1}{L}e^{\mathi (k_1-k_2)l} (-1)^{l}f_1(k_2,k_1)\;.
\end{eqnarray}

\section{Second-order matrix elements for optical excitations}
\label{appB}

\subsection{Second order in the Hubbard interaction}

As intermediate states we may have $|m\rangle=|11\rangle$ and
$|m\rangle=|21\rangle$. Due to spin symmetry, 
the contribution from 
$|m\rangle=|12\rangle$ equals that from 
$|m\rangle=|21\rangle$.
We find
\begin{eqnarray}
M_{2,(s,t)}^{U^2 |11\rangle}(e;p_1,p_2)
&=& 
\left(\frac{U^2}{2L^2}\right)\sum_{k_1,k_2,k_3,k_4 }
\frac{1}{e-\sum_{j=1}^{4}E(k_j)}
\nonumber\\
&&\bigl[\delta_{p_1,k_2} -\delta_{p_1,k_1} 
\pm \left[\delta_{p_1,k_4} -\delta_{p_1,k_3} \right]\bigr] \times
\{p_1\rightarrow p_2\}
\nonumber\\
&&\Bigl[\delta_{k_1-k_2+k_3-k_4,0}[u_2(k_1,k_2,k_3,k_4)]^2
\nonumber\\
&&\hphantom{\Bigl[}+\delta_{k_1-k_2+k_3-k_4,\pm \pi}[u_1(k_1,k_2,k_3,k_4)]^2
\Bigr]\; ,
\label{opt_u2_ztla11}
\end{eqnarray}
and 
\begin{eqnarray}
M_{2,(s,t)}^{U^2|21\rangle}(e;p_1,p_2)&\!=\!&  \left(\frac{U}{L}\right)^2
\sum_{ k_1<k_3,k_2<k_4,k_5,k_6 }\frac{1}{e-\sum_{j=1}^6 E(k_j)}
\nonumber\\
&&
\biggl[\delta_{k_1-k_2+k_3-k_4+k_5-k_6,0}
\nonumber\\
&&
\hphantom{\biggl[}
\Bigl\{
  \delta_{p_1,k_1}\delta_{k_1,k_2}u_1(k_3,k_4,k_5,k_6) \nonumber \\
&&
\hphantom{\biggl[ \Bigl\{}
+\delta_{p_1,k_3}\delta_{k_3,k_4}u_1(k_1,k_2,k_5,k_6)
\nonumber\\
&&
\hphantom{\biggl[ \Bigl\{}
-\delta_{p_1,k_1}\delta_{k_1,k_4}u_1(k_3,k_2,k_5,k_6) \nonumber \\
&& \hphantom{\biggl[ \Bigl\{}
-\delta_{p_1,k_3}\delta_{k_3,k_2}u_1(k_1,k_4,k_5,k_6)
\Bigr\}\!\times\! \Bigl\{p_1\rightarrow p_2\Bigr\}
\nonumber\\
&&\hphantom{\biggl[}
+\delta_{k_1-k_2+k_3-k_4+k_5-k_6,\pm \pi}
\nonumber\\
&&
\hphantom{\biggl[}
\Bigl\{
  \delta_{p_1,k_1}\delta_{k_1,k_2}u_2(k_3,k_4,k_5,k_6)
\nonumber \\
&&\hphantom{\biggl[ \Bigl\{}
+\delta_{p_1,k_3}\delta_{k_3,k_4}u_2(k_1,k_2,k_5,k_6)
\label{opt_u2_dtla21}\\
&&\hphantom{\biggl[ \Bigl\{}
-\delta_{p_1,k_1}\delta_{k_1,k_4}u_2(k_3,k_2,k_5,k_6)
\nonumber \\
&&\hphantom{\biggl[ \Bigl\{}
-\delta_{p_1,k_3}\delta_{k_3,k_2}u_2(k_1,k_4,k_5,k_6)\Bigr\}
\!\times\! \Bigl\{p_1\rightarrow p_2\Bigr\}\biggr]\,.
\nonumber 
\end{eqnarray}
The six-fold sum can be reduced to a three-fold sum.
Nevertheless,
expressions like this show that higher-order terms
in the perturbation theory cannot be handled numerically
because they involve five and more particle-hole excitations.

\subsection{Second order in the long-range interaction}

The particle-hole excitation in $|(s,t)\rangle$ can be destroyed.
Therefore, the state $|{\rm FS}\rangle$ with no particle-hole
excitations also contributes,
\begin{equation}
M_{2, (s,t)}^{V^2|00\rangle}(e;p_1,p_2)
= V^2 ((1\pm1)^2 v_0(p_1)v_0(p_2))/(2 e)\; .
\label{Vsquare00}
\end{equation}
Next, the states $|11\rangle$ contribute
\begin{eqnarray}
M_{2, (s,t)}^{V^2|11\rangle}(e;p_1,p_2)
&=& \frac{1}{2}\left(\frac{V}{L}\right)^2
\sum_{ k_1,k_2,k_3,k_4}\frac{1}{e-\sum_{j=1}^4 E(k_j)}
\nonumber\\
&&\biggl[  
\delta_{k_1-k_2+k_3-k_4,0}\nonumber \\
&&\hphantom{\biggl[  }\Bigl\{    
\delta_{k_1,k_2}\delta_{k_3,k_4}L\left(\delta_{p_1,k_1} v_0(k_3)  
\pm \delta_{p_1,k_3} v_0(k_1)\right)
\nonumber\\
&&\hphantom{\biggl[\Bigl\{}    
+\left(\delta_{p_1,k_2} -\delta_{p_1,k_1} \right)v_2(k_1,k_2,k_3,k_4)
\nonumber\\
&&\hphantom{\biggl[\Bigl\{}    
\pm \left(\delta_{p_1,k_4}-\delta_{p_1,k_3} \right)v_2(k_3,k_4,k_1,k_2)
\Bigr\}\nonumber \\
&& \hphantom{\biggl[} \times \Bigl\{p_1\rightarrow p_2\Bigr\}
\nonumber\\
&&\hphantom{\biggl[}
+\delta_{k_1-k_2+k_3-k_4,\pm \pi}
\nonumber\\
&&\hphantom{\biggl[}\Bigl\{\left(\delta_{p_1,k_2}-\delta_{p_1,k_1} \right)
v_1(k_1,k_2,k_3,k_4)\nonumber\\
&&\hphantom{\biggl[\Bigl\{}
\pm \left(\delta_{p_1,k_4} -\delta_{p_1,k_3} \right)v_1(k_3,k_4,k_1,k_2)
\Bigr\}\nonumber \\
&& \hphantom{\biggl[}  \times \Bigl\{p_1\rightarrow p_2\Bigr\}
\biggr]\; . \label{Vsquare11}
\end{eqnarray}
The states $|20\rangle$ and $|02\rangle$ equally contribute
\begin{eqnarray}
2 M_{2, (s,t)}^{V^2|20\rangle}(e;p_1,p_2)
&= & \left(\frac{V}{L}\right)^2
\sum_{k_1<k_3,k_2<k_4}\frac{1}{e-\sum_{j=1}^4 E(k_j)}
\nonumber\\
&&\biggl[  \delta_{k_1-k_2+k_3-k_4,0} \nonumber \\
&& \hphantom{\biggl[}
\Bigl\{\left(\delta_{k_1,k_2}\delta_{k_3,k_4}
-\delta_{k_1,k_4}\delta_{k_3,k_2}\right) \nonumber \\
&& \hphantom{\biggl[ \Bigl\{ }
\times L\left[\delta_{p_1,k_1} v_0(k_3)  + \delta_{p_1,k_3} v_0(k_1)\right]^2
\nonumber\\
&& \hphantom{\biggl[ \Bigl\{ }
\left(\delta_{p_1,k_2}-\delta_{p_1,k_1}\right) v_2(k_1,k_2,k_3,k_4)
\nonumber \\
&& \hphantom{\biggl[ \Bigl\{ }
 +\left(\delta_{p_1,k_4} -\delta_{p_1,k_3} \right)v_2(k_3,k_4,k_1,k_2)
\nonumber\\ 
&& \hphantom{\biggl[ \Bigl\{ }
-\left(\delta_{p_1,k_2} -\delta_{p_1,k_3} \right)v_2(k_1,k_4,k_3,k_2)
\nonumber \\
&& \hphantom{\biggl[ \Bigl\{ }
-\left(\delta_{p_1,k_4} -\delta_{p_1,k_1} \right)v_2(k_3,k_2,k_1,k_4)\Bigr\}
\nonumber \\
&& \hphantom{\biggl[}
\times \Bigl\{p_1 \rightarrow p_2\Bigr\}
\nonumber\\
&& \hphantom{\biggl[}
+\delta_{k_1-k_2+k_3-k_4,\pm \pi}\nonumber\\
&& \hphantom{\biggl[}
\Bigl\{ \left(\delta_{p_1,k_2} -\delta_{p_1,k_1}\right)v_1(k_1,k_2,k_3,k_4)
\nonumber \\
&& \hphantom{\biggl[ \Bigl\{ }
+\left(\delta_{p_1,k_4} -\delta_{p_1,k_3} \right)v_1(k_3,k_4,k_1,k_2)
\nonumber\\ 
&& \hphantom{\biggl[ \Bigl\{ }
-\left(\delta_{p_1,k_2} -\delta_{p_1,k_3} \right)v_1(k_1,k_4,k_3,k_2)
\nonumber \\
&&\hphantom{\biggl[ \Bigl\{ }
-\left(\delta_{p_1,k_4} -\delta_{p_1,k_1} \right)v_1(k_3,k_2,k_1,k_4)\Bigr\}
\nonumber \\
&& \hphantom{\biggl[}
\times \Bigl\{p_1\rightarrow p_2\Bigr\}\biggr]\;.
\label{Vsquare20}
\end{eqnarray}
Excitations with three particles and holes equally contribute.
We have two equal contributions from $|30\rangle$ and $|03\rangle$,
\begin{eqnarray}
2M_{2, (s,t)}^{V^2|30\rangle}(e;p_1,p_2)
&=& \left(\frac{V}{L}\right)^2
\sum_{k_1<k_3<k_5,k_2<k_4<k_6}\frac{1}{e-\sum_{j=1}^6 E(k_j)}
\nonumber\\
&&  \sum_{f=1,2} \delta_{k_1-k_2+k_3-k_4+k_5-k_6,\pm (f-1)\pi}
\nonumber \\
&& \Bigl\{
\delta_{k_5,k_6}\delta_{p_1,k_5}
\left(v_f(k_1,k_2,k_3,k_4)- (k_1\leftrightarrow k_3)\right)
\nonumber\\
&&\hphantom{\Bigl\{}
+\delta_{k_3,k_4}\delta_{p_1,k_3}
\left(v_f(k_1,k_2,k_5,k_6)- (k_1\leftrightarrow k_5)\right)
\nonumber\\
&&\hphantom{\Bigl\{}
+\delta_{k_1,k_2}\delta_{p_1,k_1}
\left(v_f(k_3,k_4,k_5,k_6)- (k_3\leftrightarrow k_5)\right)
\nonumber\\
&&\hphantom{\Bigl\{}
-\delta_{k_1,k_4}\delta_{p_1,k_1}
\left(v_f(k_3,k_2,k_5,k_6)- (k_3\leftrightarrow k_5)\right)
\nonumber\\
&&\hphantom{\Bigl\{}
-\delta_{k_1,k_6}\delta_{p_1,k_1}
\left(v_f(k_3,k_4,k_5,k_2)- (k_3\leftrightarrow k_5)\right)
\nonumber\\
&&\hphantom{\Bigl\{}
-\delta_{k_3,k_2}\delta_{p_1,k_3}
\left(v_f(k_1,k_4,k_5,k_6)- (k_1\leftrightarrow k_5)\right)
\nonumber\\
&&\hphantom{\Bigl\{}
-\delta_{k_3,k_6}\delta_{p_1,k_3}
\left(v_f(k_1,k_2,k_5,k_4)- (k_1\leftrightarrow k_5)\right)
\nonumber\\
&&\hphantom{\Bigl\{}
-\delta_{k_5,k_2}\delta_{p_1,k_5}
\left(v_f(k_1,k_6,k_3,k_4)- (k_1\leftrightarrow k_3)\right)
\nonumber\\
&&\hphantom{\Bigl\{}
-\delta_{k_5,k_4}\delta_{p_1,k_5}
\left(v_f(k_1,k_2,k_3,k_6)- (k_1\leftrightarrow k_3)\right)
\nonumber \\
&&\times \Bigl\{p_1 \rightarrow p_2\Bigr\}\; .
\label{Vsquare30}
\end{eqnarray}
Moreover, we find from the states $|21\rangle$ and $|12\rangle$,
\begin{eqnarray}
2M_{2, (s,t)}^{V^2|21\rangle}(e;p_1,p_2)
&=&\left(\frac{V}{L}\right)^2\sum_{k_1<k_3,k_2<k_4,k_5,k_6}
\frac{1}{e-\sum_{j=1}^6 E(k_j)}
\nonumber\\
&&\biggl[  \delta_{k_1-k_2+k_3-k_4+k_5-k_6,0}
\nonumber\\
&&\hphantom{\biggl[}
\Bigl\{
\delta_{k_3,k_4}\delta_{p_1,k_3} v_1(k_1,k_2,k_5,k_6) \nonumber \\
&& \hphantom{\biggl[\Bigl\{}
+\delta_{k_1,k_2}\delta_{p_1,k_1}v_1(k_3,k_4,k_5,k_6)
\nonumber\\
&& \hphantom{\biggl[\Bigl\{}
-\delta_{k_1,k_4}\delta_{p_1,k_1}v_1(k_3,k_2,k_5,k_6)\nonumber \\
&& \hphantom{\biggl[\Bigl\{}
-\delta_{k_3,k_2}\delta_{p_1,k_3}v_1(k_1,k_4,k_5,k_6)
\nonumber\\
&&\hphantom{\biggl[\Bigl\{}
\pm\delta_{k_5,k_6}\delta_{p_1,k_5}v_1(k_1,k_2,k_3,k_4) \nonumber \\
&& \hphantom{\biggl[\Bigl\{}
\mp \delta_{k_5,k_6}\delta_{p_1,k_5}v_1(k_1,k_4,k_3,k_2)\Bigr\}
\nonumber \\
&&\hphantom{\biggl[}
\times \Bigl\{p_1\rightarrow p_2\Bigr\}
\nonumber\\
&&\hphantom{\biggl[}
+ \delta_{k_1-k_2+k_3-k_4+k_5-k_6,\pm \pi}
\nonumber\\
&&\hphantom{\biggl[}
\Bigl\{
\delta_{k_3,k_4}\delta_{p_1,k_3}v_2(k_1,k_2,k_5,k_6) \nonumber \\
&&\hphantom{\biggl[\Bigl\{}
+\delta_{k_1,k_2}\delta_{p_1,k_1}v_2(k_3,k_4,k_5,k_6) \nonumber\\
&&\hphantom{\biggl[\Bigl\{}
-\delta_{k_1,k_4}\delta_{p_1,k_1}v_2(k_3,k_2,k_5,k_6) \nonumber \\
&&\hphantom{\biggl[\Bigl\{}
-\delta_{k_3,k_2}\delta_{p_1,k_3}v_2(k_1,k_4,k_5,k_6)
\nonumber\\
&&\hphantom{\biggl[\Bigl\{}
\pm
\delta_{k_5,k_6}\delta_{p_1,k_5}v_2(k_1,k_2,k_3,k_4) \nonumber \\
&&\hphantom{\biggl[\Bigl\{}
\mp \delta_{k_5,k_6}\delta_{p_1,k_5}v_2(k_1,k_4,k_3,k_2)\Bigr\}
\nonumber \\
&&\hphantom{\biggl[}
\times \Bigl\{p_1\rightarrow p_2\Bigr\}\biggr]\; .
\label{Vsquare21}
\end{eqnarray}
The $\delta$-conditions reduce the six-fold
summations to a numerically tractable problem.

\subsection{Second-order mixed interactions}

Only two terms add to the matrix elements in second order.
The states $|11\rangle$ give
\begin{eqnarray}
M_{2, (s,t)}^{UV|11\rangle}(e;p_1,p_2)
&=& \left(\frac{UV}{2L^2}\right)
\sum_{k_1,k_2,k_3,k_4}\frac{1}{e-\sum_{j=1}^4 E(k_j)}
\nonumber\\
&&\biggl(  \delta_{k_1-k_2+k_3-k_4,0}\nonumber \\
&&\hphantom{\biggl( }
\biggl[
\Bigl\{  
\left(\delta_{p_2,k_2}-\delta_{p_2,k_1} \right)u_2(k_1,k_2,k_3,k_4) \nonumber 
\\
&&
\hphantom{\biggl( \biggl[\Bigl\{  }
\pm \left(\delta_{p_2,k_4} -\delta_{p_2,k_3}\right)u_2(k_3,k_4,k_1,k_2)\Bigr\}
\nonumber\\
&&\hphantom{\biggl(\biggl[ }
\times \Bigl\{  
\left(\delta_{p_1,k_2} -\delta_{p_1,k_1}\right)v_2(k_1,k_2,k_3,k_4)
\nonumber \\
&&\hphantom{\biggl(\biggl[ \Bigl\{ \times }
\pm \left(\delta_{p_1,k_4} -\delta_{p_1,k_3}\right)v_2(k_3,k_4,k_1,k_2)\Bigr\}
\nonumber \\
&&\hphantom{\biggl(\biggl[ }
+\{p_1 \leftrightarrow p_2\}\biggr]\nonumber\\
&&\hphantom{\biggl(  }
+\delta_{k_1-k_2+k_3-k_4,\pm \pi} \nonumber \\
&&\hphantom{\biggl(  }
\biggl[ 
\Bigl\{  \left(\delta_{p_2,k_2}-\delta_{p_2,k_1} \right)u_1(k_1,k_2,k_3,k_4)
 \nonumber\\
&& \hphantom{\biggl(\biggl[ \Bigl\{} 
\pm \left(\delta_{p_2,k_4} -\delta_{p_2,k_3}\right)u_1(k_3,k_4,k_1,k_2)\Bigr\}
\nonumber\\
&& \hphantom{\biggl( \biggl[ } 
\times 
\Bigl\{ \left(\delta_{p_1,k_2} -\delta_{p_1,k_1}\right)v_1(k_1,k_2,k_3,k_4)
\nonumber \\
&& \hphantom{\biggl(\biggl[ \Bigl\{\times} 
\pm \left(\delta_{p_1,k_4} -\delta_{p_1,k_3}\right)v_1(k_3,k_4,k_1,k_2)\Bigr\}
\nonumber \\
&& \hphantom{\biggl(\biggl[ } 
+\{p_1 \leftrightarrow p_2\}\biggr]\biggr)\; . \label{UV11}
\end{eqnarray}
The states~$|21\rangle$ and $|12\rangle$ equally contribute and give
\begin{eqnarray}
2M_{2, (s,t)}^{UV|21\rangle}(e;p_1,p_2)
&=& \left(\frac{UV}{L^2}\right)
\sum_{k_1<k_3,k_2<k_4,k_5,k_6}\frac{1}{e-\sum_{j=1}^6 E(k_j)}\nonumber \\
&&\sum_{f=1,2} \delta_{k_1-k_2+k_3-k_4+k_5-k_6,\pm (f-1)\pi}
\nonumber\\
&&\biggl[\Bigl\{\delta_{k_3,k_4}\delta_{p_2,k_3}u_f(k_1,k_2,k_5,k_6) 
\nonumber\\
&&\hphantom{\biggl[\Bigl\{}
+\delta_{k_1,k_2}\delta_{p_2,k_1}u_f(k_3,k_4,k_5,k_6)
\nonumber\\
&&\hphantom{\biggl[\Bigl\{}
-\delta_{k_1,k_4}\delta_{p_2,k_1}u_f(k_3,k_2,k_5,k_6)\nonumber\\
&&\hphantom{\biggl[\Bigl\{}
-\delta_{k_3,k_2}\delta_{p_2,k_3}u_f(k_1,k_4,k_5,k_6)\Bigr\}
\nonumber\\
&&\hphantom{\biggl[}
\times \Bigl\{\delta_{k_3,k_4}\delta_{p_1,k_3}v_f(k_1,k_2,k_5,k_6)\nonumber\\
&&\hphantom{\biggl[\Bigl\{\times}
+\delta_{k_1,k_2}\delta_{p_1,k_1}v_f(k_3,k_4,k_5,k_6)\nonumber \\
&&\hphantom{\biggl[\Bigl\{\times}
-\delta_{k_1,k_4}\delta_{p_1,k_1}v_f(k_3,k_2,k_5,k_6)\nonumber \\
&&\hphantom{\biggl[\Bigl\{ \times}
-\delta_{k_3,k_2}\delta_{p_1,k_3}v_f(k_1,k_4,k_5,k_6) \nonumber \\
&&\hphantom{\biggl[\Bigl\{ \times}
\pm(\delta_{k_5,k_6}\delta_{p_1,k_5}v_f(k_1,k_2,k_3,k_4) \nonumber \\
&&\hphantom{\biggl[\Bigl\{ \times}
-\delta_{k_5,k_6}\delta_{p_1,k_5}v_f(k_1,k_4,k_3,k_2))\Bigr\} \nonumber \\
&&\hphantom{\biggl[}
+\Bigr\{p_1 \leftrightarrow p_2\Bigr\}\biggl]\; .
\label{UV21}
\end{eqnarray}

\end{document}